\documentclass[%
reprint,
amsmath,amssymb,
aps,longbibliography
]{revtex4-1}
\usepackage{url}
\usepackage[colorlinks,linkcolor=blue]{hyperref}
\usepackage{graphicx}
\usepackage{dcolumn}
\usepackage{bm}

\begin{document}
	\title{Orbital angular momentum-enhanced phase estimation using non-Gaussian state with photon loss}
	\author{Yong-Jian Chen$^{1}$}
	
	\author{Jin-Wei Gao$^{2}$}\email[]{jwgao@cwnu.edu.cn}
	\author{Jin-Xuan Han$^{1}$}
	\author{Zhong-Hui Yuan$^{1}$}
	\author{Ruo-Qi Li$^{1}$}
	\author{Yong-Yuan Jiang$^{1,3,4,5}$}
	\author{Jie Song$^{1,3,4,5}$}\email[]{jsong@hit.edu.cn}
	
	\affiliation{$^{1}$School of Physics, Harbin Institute of Technology, Harbin 150001, China\\
		$^{2}$School of Electronic Information Engineering, China West Normal University, Nanchong 637002, Sichuan, China \\		
		$^{3}$Key Laboratory of Micro-Nano Optoelectronic Information System, Ministry of Industry and Information Technology, Harbin 150001, China\\
		$^{4}$Key Laboratory of Micro-Optics and Photonic Technology of Heilongjiang Province, Harbin Institute of Technology, Harbin 150001, China\\
		$^{5}$Collaborative Innovation Center of Extreme Optics, Shanxi University, Taiyuan, Shanxi 030006, People's Republic of China
	}
	
	\begin{abstract}
		
		This study investigates the use of orbital angular momentum (OAM) to enhance phase estimation in Mach-Zehnder interferometers (MZIs) by employing non-Gaussian states as input resources in the presence of noise.  Our research demonstrates that non-Gaussian states, particularly the photon-subtraction-then-addition (PSA) state, exhibit the best sensitivity in the presence of symmetric noise. Additionally, higher-order of Bose operator of non-Gaussian states provide better sensitivity for symmetric noise. OAM can mitigate the deterioration of noise, making it possible to estimate small phase shifts $\theta \rightarrow 0$. OAM enhances the resolution and sensitivity of all input states and mitigating the deterioration caused by photon loss. Additionally, OAM enhances the resolution and sensitivity of all input states, enabling the sensitivity to approach the $1/N$ limit even under significant photon loss (e.g.,$50\%$ symmetric photon loss). These results hold promise for enhancing the sensitivity and robustness of quantum metrology, particularly in the presence of significant photon loss.

	\end{abstract}
	\maketitle
	\section{Introduction}

	Interferometric phase estimation is a critical research topic in various growing fields, such as gravitational wave detection~\cite{2019MTse} and quantum-enhanced dark matter searches~\cite{Backes2021}. Among optical interferometers, Mach-Zehnder Interferometer is a widely used and practical tool for estimating small phase changes~\cite{Hong2021, PhysRevLett.128.040504, PhysRevA.104.033521, PhysRevA.104.013725, PhysRevA.105.033704, PhysRevApplied.17.024061, PhysRevA.103.032617}. Improving the sensitivity of phase estimation in the MZI has been the focus of extensive research in recent years~\cite{ge2018distributed, zhang2021improved, Pang2017, WANG2023, 2022guo, 2021zhong, 2022kumar}. Many efforts have been made to find better input state candidates to improve the resolution and sensitivity. With a classical light field as an input resource for the MZI, sensitivity can only reach the shot-noise limit (SNL).However, non-classical states of light provide numerous potential choices for quantum metrology. The Heisenberg limit (HL) is theoretically attainable with the use of a non-classical light state~\cite{2019PRL_non_class}. The two-mode squeezing vacuum state (TMSV) with parity measurement beats the HL while saturating the quantum Cramér–Rao Bound (QCRB)~\cite{2010PRL}.
	
	Non-Gaussian states~\cite{2020PRXgenerate, 2021PRAgenerate, 2022PRLnon-gaussian_qubit, PRXQuantum.2.030204, 2022PRapplied_non-gaussian,QI2018}, which have emerged as promising candidates, offer the potential to enhance quantum metrology~\cite{2021PRA_QKD, 2020zubairy_QKD, PhysRevASingh, 2021Bose_quantumtele, 2021PRL_nongaussian_sensi}. A variety of non-classical states, including photon-subtracted and photon-added states~\cite{ouyang2016quantum, wang2019quantum}, as well as coherent, squeezed vacuum states and N00N-like states~\cite{ataman2018phase, PhysRevA.105.023718, PhysRevA.105.012606}, have been studied to improve phase estimation. From a practical standpoint, the impact of noise~\cite{Thekkadath2020, gatto2022heisenberg, oh2020optimal, frascella2021overcoming}, particularly photon loss-induced decoherence, poses a significant challenge in the pursuit of enhanced sensitivity and resolution~\cite{PRXQuantum.3.010202, PhysRevA.105.013704}. Both theoretical and experimental research have indicated that Orbital Angular Momentum (OAM) offers a degree of resilience to light beams in noisy environments~\cite{2021PRLN00N, LiuDEC24, nape2021vector}. As such, the exploration of non-Gaussian states and OAM becomes an important area of study. These techniques harbor the potential to facilitate precise phase estimation and heightened sensitivity

	Our investigation highlights the potential of utilizing OAM to increase the sensitivity of phase estimation in the MZI system, by employing non-Gaussian states as input states. We use balanced (50:50) beam splitter MZI and parity detection as the optimal setup~\cite{2018PRA_MZI, 2019pra_MZI, 2021PRA_parity, 2022pra_parity}. The results of our research reveal that non-Gaussian states, display superior sensitivity compared to the TMSV state under symmetric and weak-symmetric noise. To further enhance sensitivity in the presence of photon loss for a specific non-Gaussian state, we employ the lower-order of the Bose operator and implement balanced photon loss. OAM effectively mitigates noise degradation, facilitating small phase shift estimation. Furthermore, OAM enhances the resolution and sensitivity of all input states while counteracting the negative impact of photon loss, and enables sensitivity to approach the $1/N$ limit even under significant photon loss, such as $50\%$ symmetric photon loss. We also analyzed the statistical properties of quantum states in the MZI system by von Neumann entropy and the Wigner function~\cite{PhysRevA.102.042405, PhysRevA.102.032413, PhysRevD.106.025002, PhysRevA.104.032423}. Therefore, this work provides prospects for realizing higher sensitivity in quantum metrology with the presence of significant noise.

	\section{\label{S2}PHASE ESTIMATION PROTOCOL ENHANCED BY NON-GAUSSIAN STATE and OAM} \label{S2}
	\subsection{The design of the MZI system and the expression of non-Gaussian state}\label{S2A}
	
	Using non-Gaussian states is a feasible method to find better input state candidates~\cite{Namekata2010,Liyun2017,quantumscissor,PRLnonclass}. Exploring PS, PA, and their superposition state as input resource gives a fine example~\cite{2019Guo,2018zhao,2018ma}. We give the expression of such states by harnessing Bose operator on the TMSV state, represented by $\hat{a}$ $ (\hat{a}^{\dagger}) | \Psi \rangle_{\rm TMSV}$. Here, we introduce a kind of non-Gaussian state determined by different sequences and order of Bose operators as
	\begin{eqnarray}\label{e1}
		&{|\hat{\Psi}\rangle}_{\rm PA}=& N_{p} {\hat{a}^{\dagger G}}{\hat{b}^{\dagger H}}  \hat{S}(\xi)   {\left|0,0\right>},\nonumber \\
		&{|\hat{\Psi}\rangle}_{\rm PS}=& N_{p} {\hat{a}^{J}}{\hat{b}^{K}} \hat{S}(\xi){\left|0,0\right>},\nonumber \\		
		&{|\hat{\Psi}\rangle}_{\rm PAS}=& N_{p} {\hat{a}^{\dagger G}}{\hat{b}^{\dagger H}}{\hat{a}^{J}}{\hat{b}^{K}} \hat{S}(\xi){\left|0,0\right>},\nonumber \\ 
		&{|\hat{\Psi}\rangle}_{\rm PSA}= & N_{p} {\hat{a}^{\dagger G}}{\hat{b}^{\dagger H}}{\hat{a}^{J}}{\hat{b}^{K}} \hat{S}(\xi){\left|0,0\right>},
	\end{eqnarray}
	where $N_{p}$ denotes a normalized parameter, $\hat{S}(\xi)$ represent two mode squeezing operator, $\xi=re^{i\psi}$, and r is know as the squeezing parameter and $\psi$ as squeezing angle, $\hat{a}$ $(\hat{a}^{\dagger})$ and $\hat{b}$ $ (\hat{b}^{\dagger})$ serve as annihilation (creation) operators for modes $a$ and $b$, respectively. Eq. (\ref{e1}) provides the formulation for the photon-addition (PA), photon-subtraction (PS), photon-addition-then-subtraction (PAS), and photon-subtraction-then-addition (PSA) states. It should be noted that $G$ and $H$ symbolize the order of the Bose operator, as discussed in previous works~\cite{2007VR,lee2012second}. Non-Gaussian states where $G=H$ exhibit superior sensitivity compared to those where $G\neq H$. Consequently, our discussion is primarily focused on the precondition of $G=H$. For $G=H=1$, we label the non-Gaussian states as PA11, PS11, PAS11, and PSA11. When $G=H=2$, these states are referred to as PA22, PS22, PAS22, and PSA22. The appended numbers in each case correspond to the values of $G$ and $H$.

\begin{figure}
		\centering
		\includegraphics[scale=0.45]{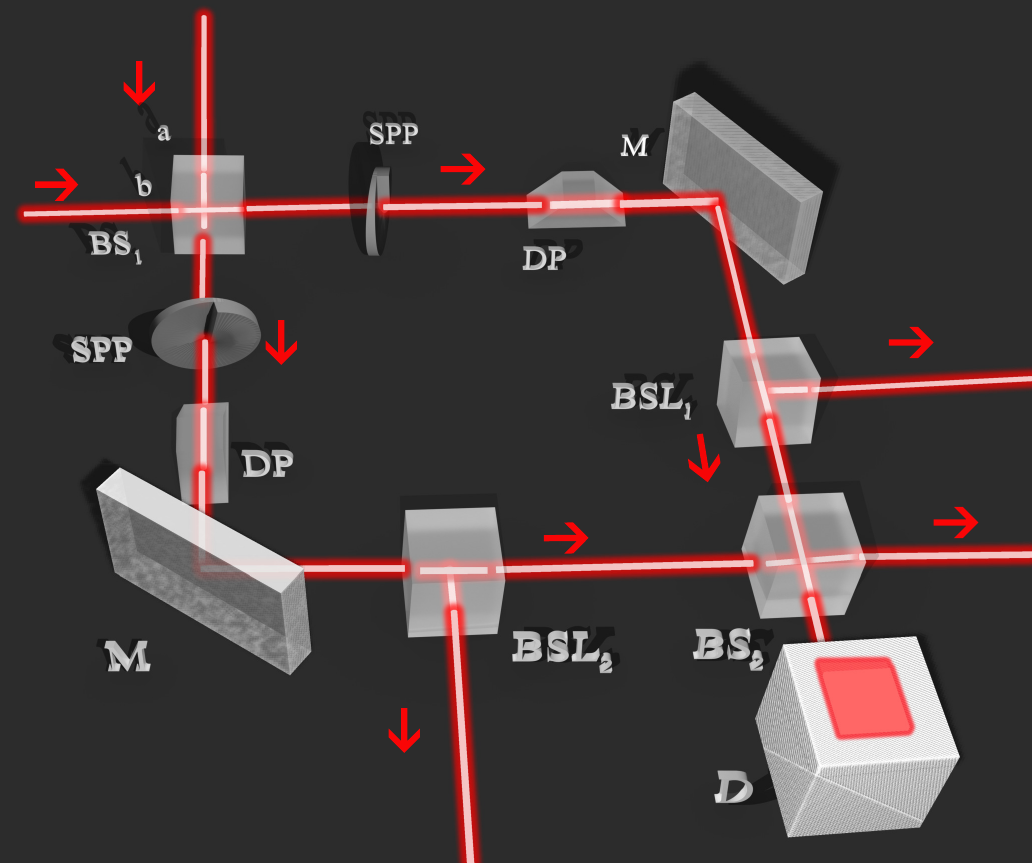}
		\caption{Schematic of Mach-Zehnder interferometer system. Fed by quantum states input from ports a and b. The light is transmitted through paths c and d and finally detected in ports e or f with Detecter (D). After a phase shift generated by Dove Prism (DP), we use two arbitrary rates Beam Splitters to simulate the photon loss (BSL). Spiral phase plates (SPP) are used to bring OAM. Mirror (M) can change the direction of the light field.}
		\label{f1}
	\end{figure}

	Fig.~\ref{f1} explicitly shows the phase estimate protocol. Two 50:50 beam splitters (BS) and two phase shifters compose the main structure of MZI. The two BSs are represented by the operators $U_{\rm BS_{1}} = \exp \left[-i \frac{\pi}{4}(\hat{a}^\dagger \hat{b}+\hat{a} \hat{b}^\dagger) \right]$ and $U_{\rm BS_{2}} = \exp \left[i \frac{\pi}{4}(\hat{a}^\dagger \hat{b}+\hat{a} \hat{b}^\dagger) \right]$, respectively. Dove Prisms (DP) embedded in modes c and d are phase shifters denoted by $\exp (-i\theta \hat{n})$, where $\hat{n}$ is the photon number operator, and $\theta$ represents the phase shift angle. The protocol proposes two opposite phase shifts $\frac{\theta}{2}$ in modes c and d. To investigate the photon loss, we insert two more BSs with transmittance $T_{\rm a(b)}$ between $\rm BS_{1}$ and $\rm BS_{2}$ to generate photon loss. 
	
	The OAM beam is characterized by a light beam that exhibits a helical phase structure within its wavefront, with each photon carrying an OAM value of $L\hbar$, where $L$ signifies the azimuthal angular parameter, also known as the topological charge of the OAM beam. As the value of $L$ increases, both the radius of the beam's cross-sectional area and the angle between the beam's Poynting vector and the optical axis increase. The topological charge $L$ (i.e., $L, 2L, 3L, \ldots$) serves to enhance the parity measurement results and sensitivity, (refer to Eq.~(\ref{e3}), Eq.~(\ref{eA9}), and Eq.~(\ref{eA11}) for further details). In this study, modulate the beam's phase (e.g. using SPP), directly imparting a helical phase onto the beam to generate an OAM beam. The OAM in the input state can act as a "gear" to magnify the phase shift term, thereby enabling the estimation of small phase shifts~\cite{OAM1, OAM2, OAM3, OAM4, OAM5}.
	
	 The light detection scheme at the output mode(s) of MZI determines the best sensitivity we can obtain. Among many detection approaches, parity measurement can exploit the potential of non-classical light and reach the QCRB~\cite{2021PRA_parity, 2017parity, 2022pra_parity} (detail see Appendix.~\ref{a1}). This work uses the parity operator in one of the output modes, represented by $ \Pi_{b}=(-1)^{\hat{b}^{\dagger} \hat{b}}$. The expectation value of the measurement signal can be represented by 
	\begin{equation}\label{e2}
		\left< \Pi_{b} \right> =\rm Tr[\rho_{\rm out} \Pi_{b}],
	\end{equation}
	where $\rho_{\rm out}$ denotes the density matrix of the output state.

	As an example, we give the expression of parity signal derived by Eq.~(\ref{e2}) using PSA11 as the input state, 
	\begin{eqnarray}\label{e3}
		&\left<  \Pi \right>_{\rm PSA11 (\theta + \pi /2)}=&\{ (1-z^{2})[-4z^{2}(2-63z^{4}+39z^{8}\nonumber\\&&+14z^{12})\rm cos\emph L(\pi+2\theta)+\emph z^{4}(-4\nonumber\\&&+315z^{4}-252z^{8}+8z{12}+4*(15\nonumber\\&&-41z^{4}+9z^{8})\rm cos2\emph L(\pi+2\theta)+4\emph z^{2}\nonumber\\&&(-9+z^{4})\rm cos3\emph L(\pi+2\theta) +z^{4}\nonumber\\&&\rm cos4\emph L(\pi+2\theta)) ] \} / \{ 8(1+\emph z^{4}+2\emph z^{2}\nonumber\\&&\rm cos2\emph L\theta)^{\frac{1}{2}} (1+\emph z^{4}\nonumber\\&&+2\emph z^{2}\rm cos\emph L(\pi+2\theta))      \}, 
	\end{eqnarray}	
	where $z$ denotes $\rm tanh(\emph r)$, and $\theta$ is the phase shift. The detail of the calculation and the result of other input states are shown in Appendix~\ref{a2}. As shown in Eq.~(\ref{e3}), there are three parameters that determine $\left<  \Pi \right>$: squeezing parameter $r$, phase shift $\theta$, and OAM quantum number $L$.

	\subsection{The enhancement brought by non-Gaussian state and  OAM}\label{S2A}

\begin{figure}[htb]
		\centering
		\includegraphics[scale=0.3]{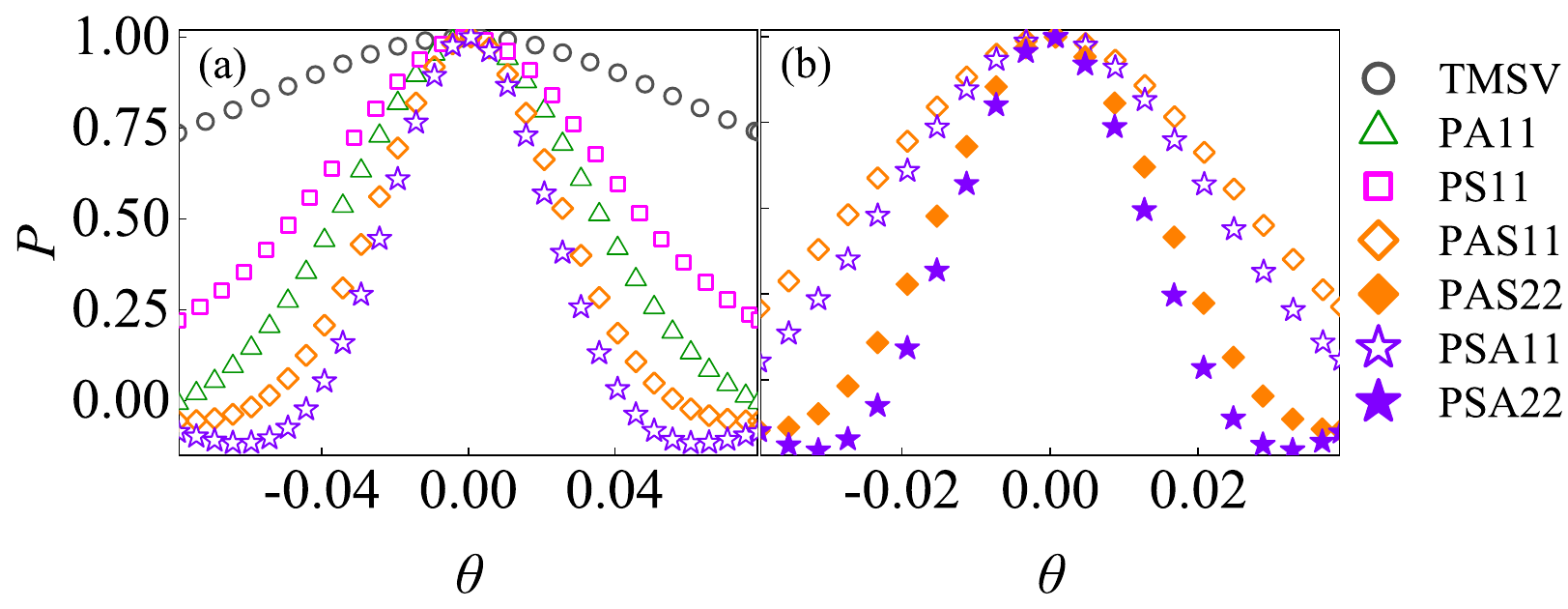}
		\caption{Normalized detection signal $P$ as a function of phase shift $\theta$ for all input states with squeezing parameter $r=1.096$. The black circle denotes the TMSV state. As for the non-Gaussian states, we use PA, PS, PAS, and PSA (triangles, squares, diamonds, and stars). We use hollow icons to represent $G=H=1$ and solid icons for $G=H=2$.}
		\label{f2}
	\end{figure}

	We plot Fig.~\ref{f2} to show the signal of the parity detection $\left< \Pi \right>$ as a function of phase shift $\theta$, where the label $P$ is used to represent $\left< \Pi \right>$. The full width at half-maximum (FWHM) of the signal curve is one of the universal criteria for determining resolution. For a narrower signal peak, a small phase shift leads to a distinct change in the $P$ value, indicating a higher resolution. It is apparent from Fig.~\ref{f2}(a) that for the same squeezing parameter, TMSV has worse, and PSA11 owns the best resolution. Fig.~\ref{f2}(b) demonstrates that the higher-order Bose operator brings superior resolution.   

	There are only two outcomes for parity detection:  $+$ for even and $-$ for odd. According to Ref.~\cite{2013QCRBsensi}, the classical Fisher information ($F_{\rm C}$) determined by $\theta$ can be used to calculate the detection sensitivity, represented by Eq.~(\ref{e4}),
	\begin{equation}\label{e4}
		\Delta \theta=\frac{1}{\sqrt{F_{\rm C}}}=\frac{\sqrt{1-\langle \Pi_{b} \rangle}}{\left| \partial \langle \Pi_{b} \rangle / \partial \theta  \right|}.
	\end{equation}
	
	\begin{figure}[htb]
		\centering
		\includegraphics[scale=0.3]{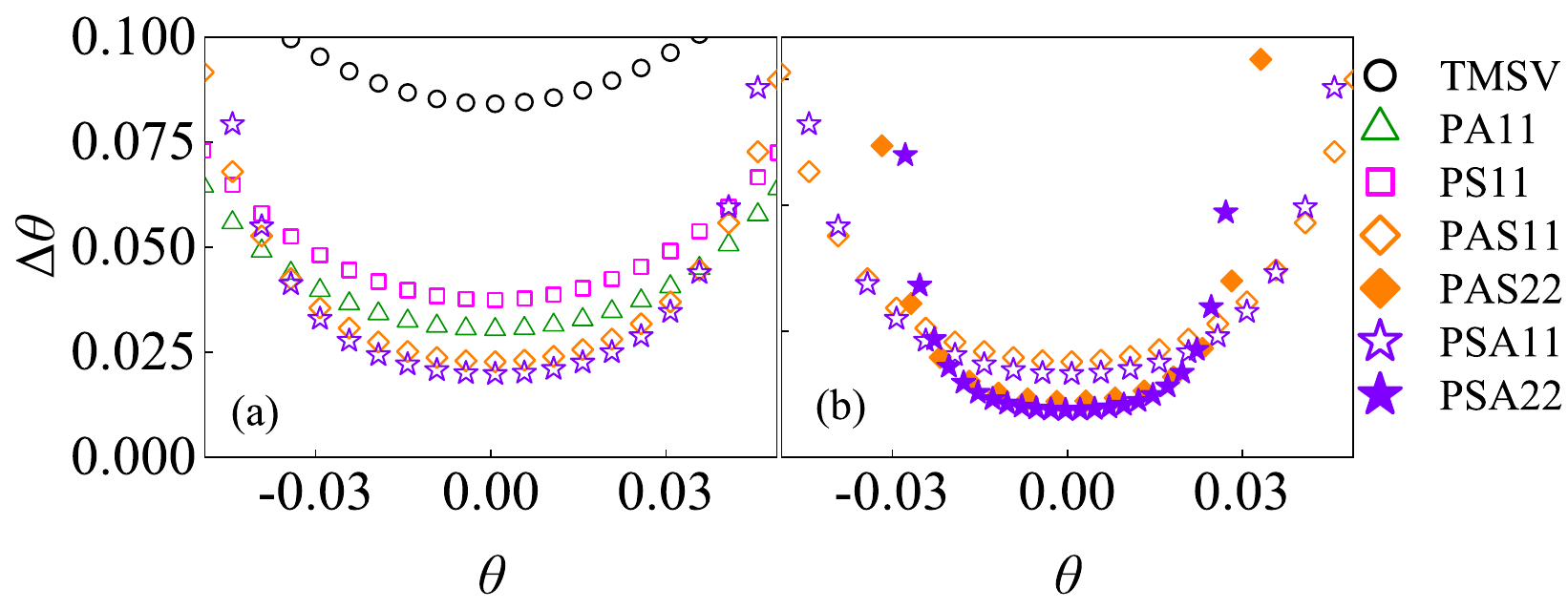}
		\caption{Phase estimate sensitivity $\Delta \theta$ as a function of the phase shift $\theta$ with squeezing parameter $r=1.096$ for all input states. Panel (a) shows the states with $G=H=1$. In panel (b), we depict the PAS and PSA state with different orders of the Bose operator. We use hollow icons to denote input states with $G=H=1$ and solid icons for $G=H=2$.}
		\label{f3}
	\end{figure}

	To investigate how phase sensitivity changes as a function of $\theta$, we plot Fig.~\ref{f3} using Eq.~(\ref{e4}). We can see that in the case of $\theta \rightarrow 0$, the lowest bound of phase sensitivity is obtained~\cite{SeshadreesanAUG24, 2013QCRBsensi}. The data in panel (a) suggests that the TMSV state has the worst, and the PSA11 state owns the best sensitivity. Panel (b) implies that the higher-order Bose operator brings a lower $\Delta \theta$ bound. Overall, these results show that with fixed squeezing parameter $r$, the PSA state owns the best resolution and sensitivity for small phase shift $10^{-3}$. Also, with the increase of $G$ and $H$, both resolution and sensitivity increase.
		
	On the other hand, we compair the sensitivity of input state as the function of mean photon number $N$ in Fig.4. As we can see from the figure, with fixed mean photon number, the TMSV state exhibits better sensitivity than non-Gaussian states when there is no photon loss. 
	
		\begin{figure}[htb]
		\centering
		\includegraphics[scale=0.29]{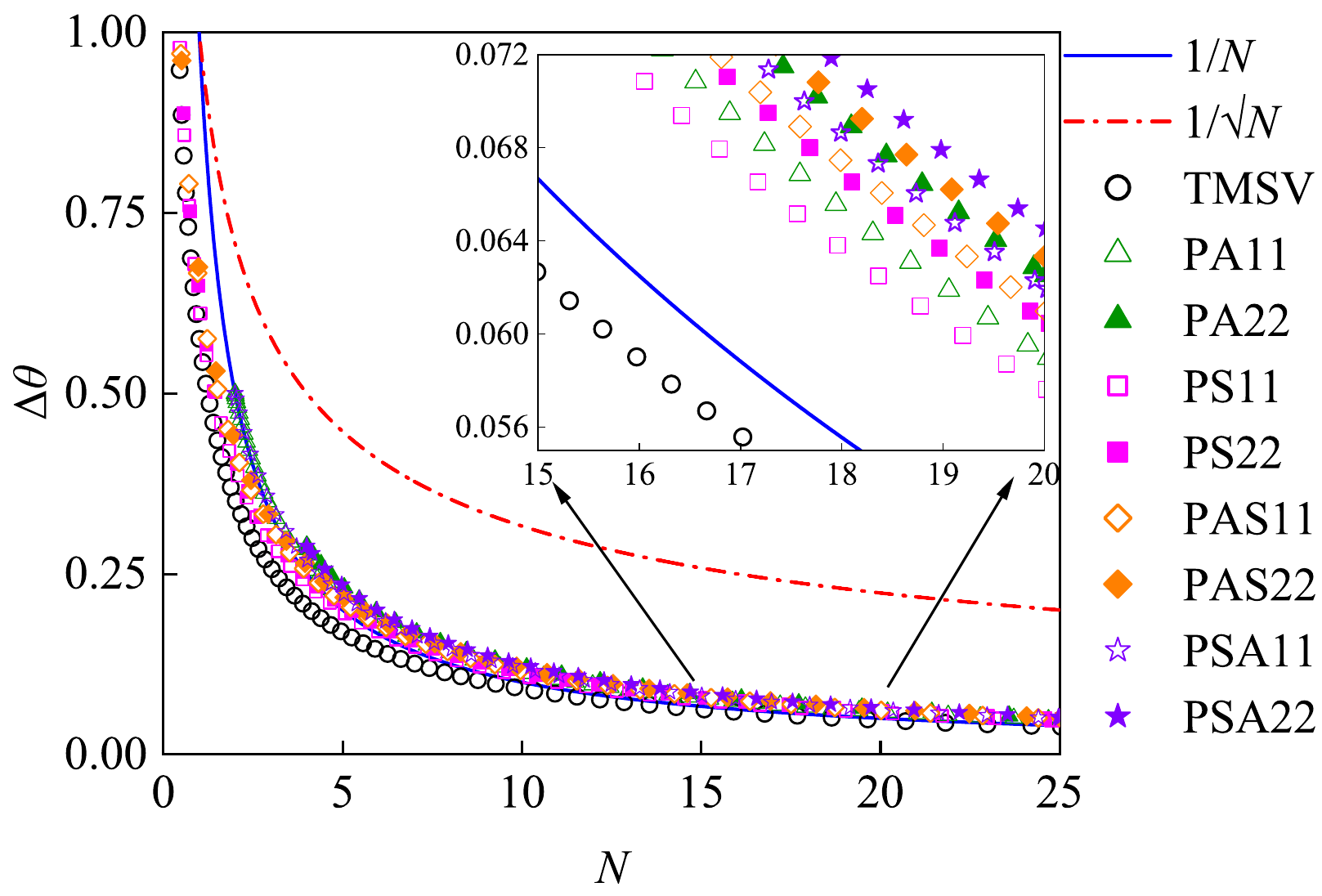}
		\caption{Sensitivity $\delta \theta$ as a function of mean photon number $N$ with no photon loss. Where the phase shift has been taken as $\theta = 10^{-4}$. The hallow (solid) icon represent the state with $G=H=1$ ($G=H=2$).} 
		\label{f4}
	\end{figure}
	
\begin{figure}[htb]
		\centering
		\includegraphics[scale=0.265]{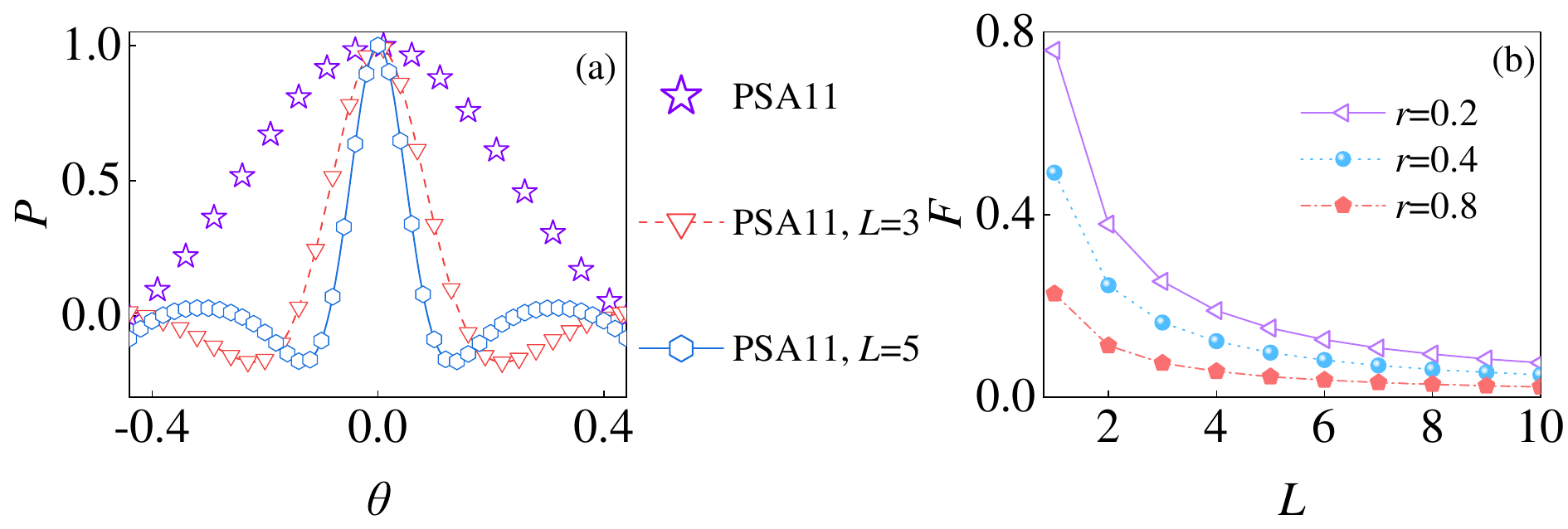}
		\caption{\textcolor{blue}{(a) Detection signal $P$ as a function of phase shift for PSA11 input state with $r=1.096$. Stars, triangles and hexagons represent the PSA11 state with non-OAM-added, $L=3$ and $L=5$, respectively. (b) FWHM of the PSA11 state varies with $L$ for $r=0.2,0.4$ and $0.8$. }}
		\label{f5}
	\end{figure}

	We also plot Fig.~\ref{f5} to visualize the enhancement of resolution brought by OAM. The figure depicts the normalized detection signal $P$ as a function of $\theta$ with various topological charge $L$. We use the signal's FWHM (denoted by the label $F$) as the criterion to compare the resolution. From Fig.~\ref{f4}(a), the results demonstrate  that the higher OAM provides superior resolution. We plot $F$ versus OAM number $L$ in panel (b) at $r=0.2,0.4$, and $0.8$. The enhancement brought by OAM is prominent, which will render higher resolution and sensitivity.

	\begin{figure}[h!tb]
		\renewcommand{\thefigure}{6}
		\centering
		\includegraphics[scale=0.39]{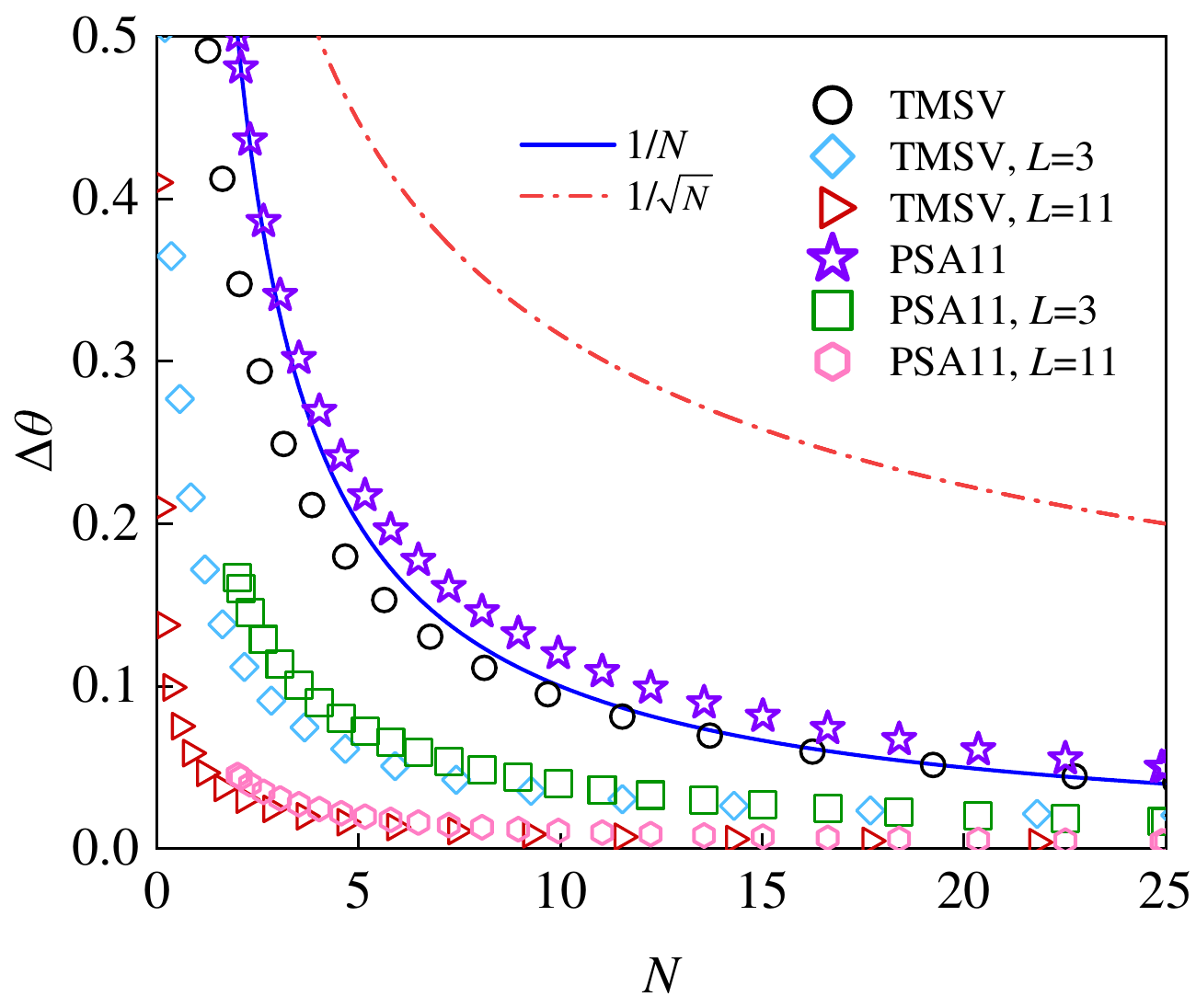}
		\caption{\textcolor{blue}{Phase estimation sensitivity $\Delta \theta$ as a function of mean photon number $N$ without noise for no-OAM-added, $L=3$ and $L=11$. The solid blue line and red dash-doted line represent $1/N$ and $1/\sqrt{N}$ limits. Star, square and hexagon represent sensitivity of the PSA11 state with non-OAM-enhancement, $L=3$, and $L=11$. Circle, diamond and triangle indicate TMSV state with non-OAM-enhancement, $L=3$ and $L=11$, respectively. The phase shift is fixed at $\theta=10^{-4}$.}}
		\label{f6}
	\end{figure}
	
	We plot the OAM-enhanced sensitivity $\Delta \theta$ against the mean photon number $N$ in Fig.~\ref{f6}, where the phase shift is fixed at $\theta=10^{-3}$. This figure uses TMSV and PSA11 state as examples to demonstrate the $\Delta \theta$ against $N$ for no-OAM-added, $L=3$ and $L=11$. Without the photon loss, the TMSV state owns better sensitivity than non-Gaussian states. Sensitivity improves with increasing $L$. 
	
	\section{\label{S3}The effect of various noise and the mitigation of it}
	
	\subsection{The influence of symmetric and asymmetric noise}\label{S3A}

	\begin{figure*}[htbp]\centering
		\renewcommand{\thefigure}{7}
		\centering
		\includegraphics[width=0.85\linewidth]{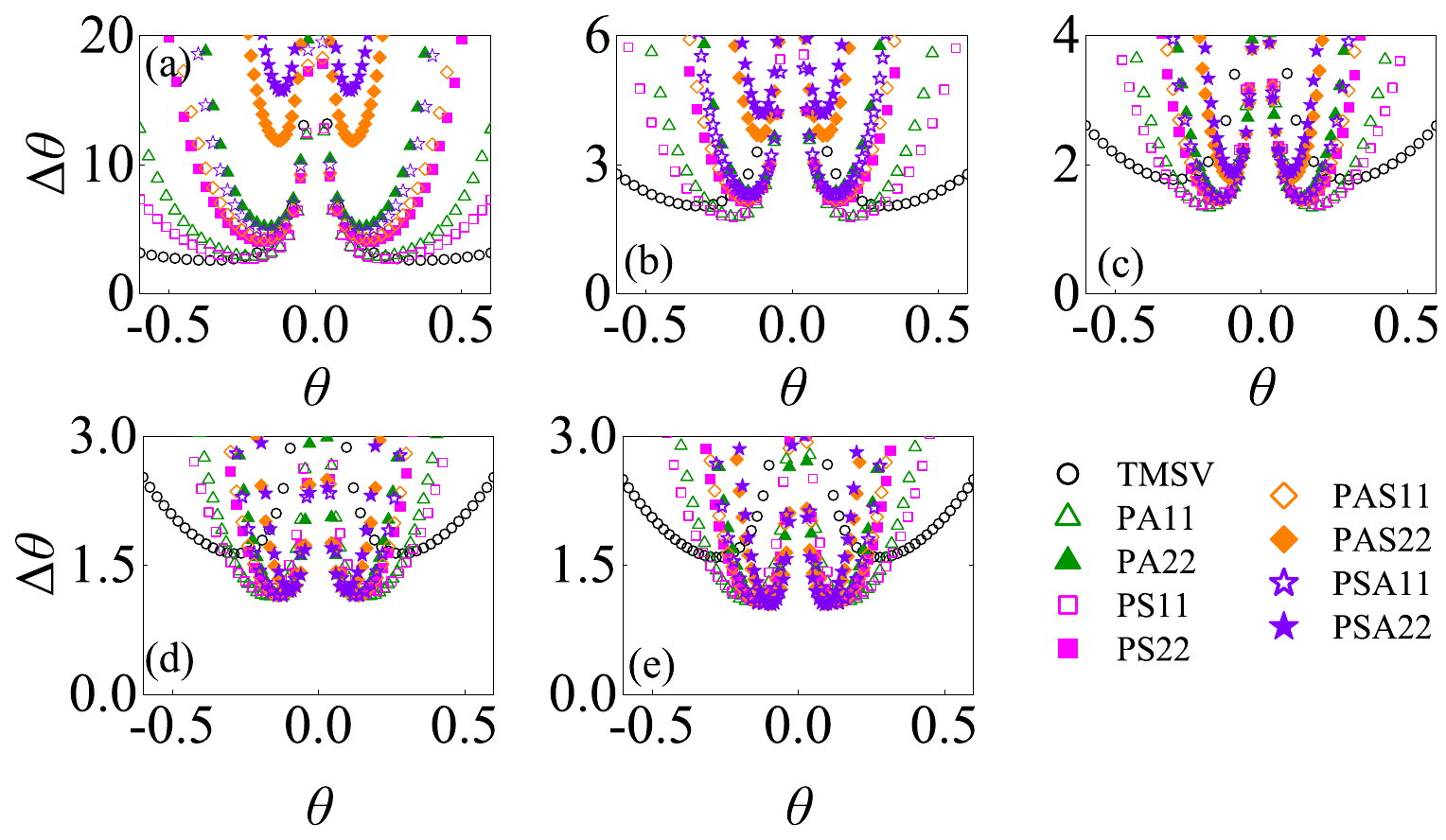}
		\caption{Phase estimate sensitivity as a function of phase shift with various photon loss. Panel (a), (b), (c), (d) and (e) represent $T_{\rm a(\rm b)}=0.1(0.9)$, $T_{\rm a(\rm b)}=0.2(0.8)$, $T_{\rm a(\rm b)}=0.3(0.7)$, $T_{\rm a(\rm b)}=0.4(0.6)$, and $T_{\rm a(\rm b)}=0.5(0.5)$, respectively. The squeezing parameter is taken as $r=1.096$. }
		\label{f7}
	\end{figure*}
	
	 The unavoidable photon loss leads to a worse  $\Delta\theta$. In the protocol, a BS is used to cause photon loss in both arms of MZI. We define $T_{\rm a}=T_{\rm b}=T$ as symmetric noise and $T_{\rm a}\neq T_{\rm b}$ as asymmetric noise, respectively. Weak-symmetric noise is defined as $T_{\rm b}-T_{\rm a} \leq 0.4$, and strong-symmetric noise is defined as $T_{\rm b}-T_{\rm a} \geq 0.4$. We investigate how the noise influences the phase estimation. We plot $\Delta \theta$ as a function of $\theta$ with various noises for all input states in Fig.~\ref{f6}, where we set the squeezing parameter as $r=1.096$. The figure reveals that the lowest bound of $\Delta \theta$ for non-Gaussian states is getting worse with the increase of $T_{\rm b}-T_{\rm a}$, as evident from panels (e) to (a). The results demonstrate that in order to further improve the sensitivity, balancing the photon loss for non-Gaussian states is an effective way. The PSA state exhibits the best sensitivity for small phase shifts in the presence of symmetric noise, while higher-order Bose operators providing even greater sensitivity~\cite{symmetricnoise}.

	\begin{figure}[htb]
		\centering
		\includegraphics[scale=0.35]{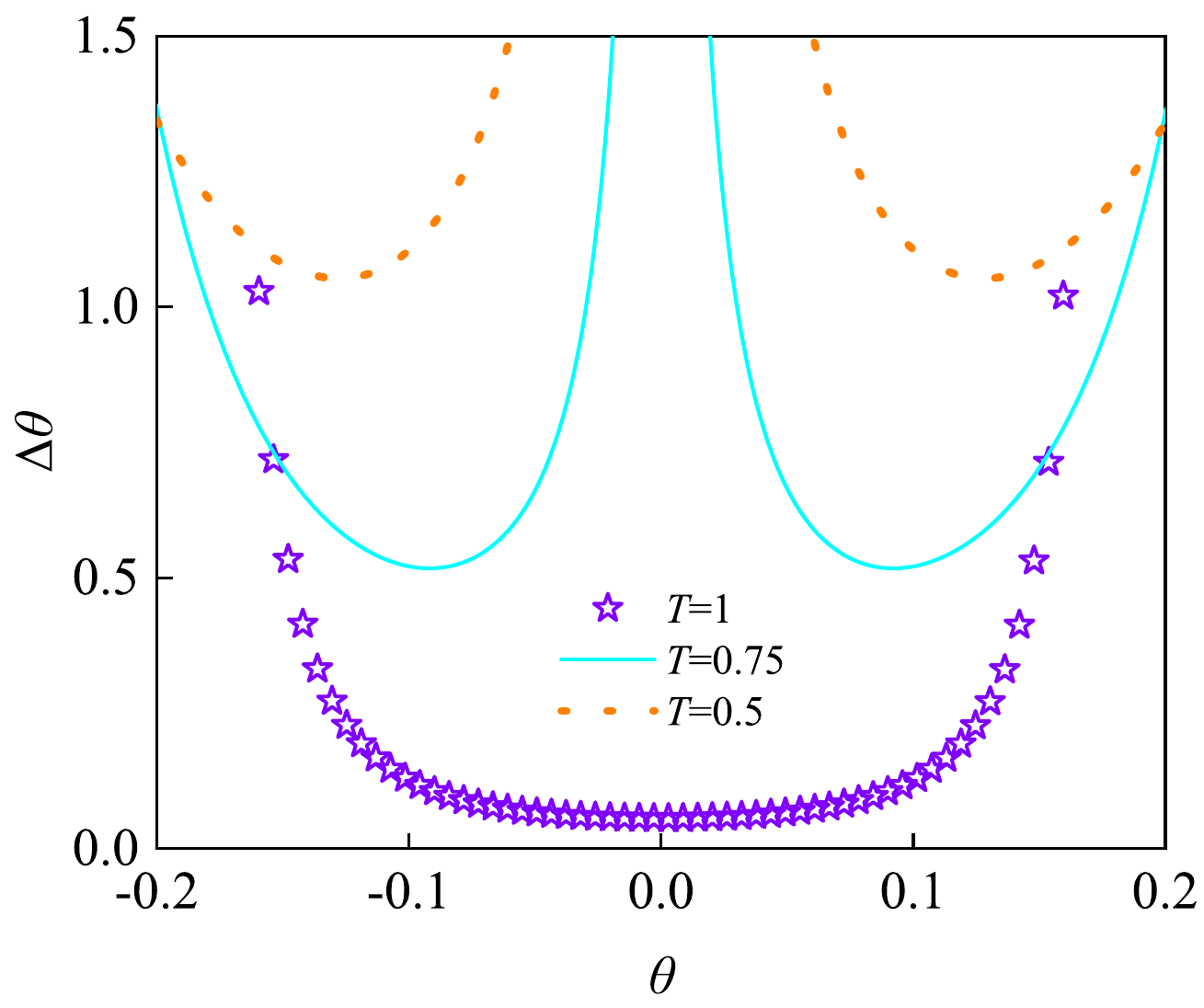}
		\caption{ The sensitivity as a function of $\theta$ for the PSA11 state, the squeezing parameter has been taken as $r=1.096$. Stars denote the $\Delta \theta$ without the photon loss. The solid line and dots represent $\Delta \theta$ for $T=0.75 $ and $0.5$,respectively.  }
		\label{f8}
	\end{figure}	
	Fig.~\ref{f8} provides significant insight into $\Delta \theta$ as a function of $\theta$ with different transmittance $T$ to investigate the influence of photon loss. We can see from Fig.~\ref{f8} that increasing photon loss leads to a higher bound of $\Delta \theta$, which means worse sensitivity. Compared to Fig.~\ref{f3}, the photon loss deteriorates $\Delta \theta$ and generates a ``gap" for $\theta \rightarrow 0$. The broader ``gap" makes estimating a small phase shift challenging. 
	\begin{figure}[htb]
		\centering
		\includegraphics[scale=0.39]{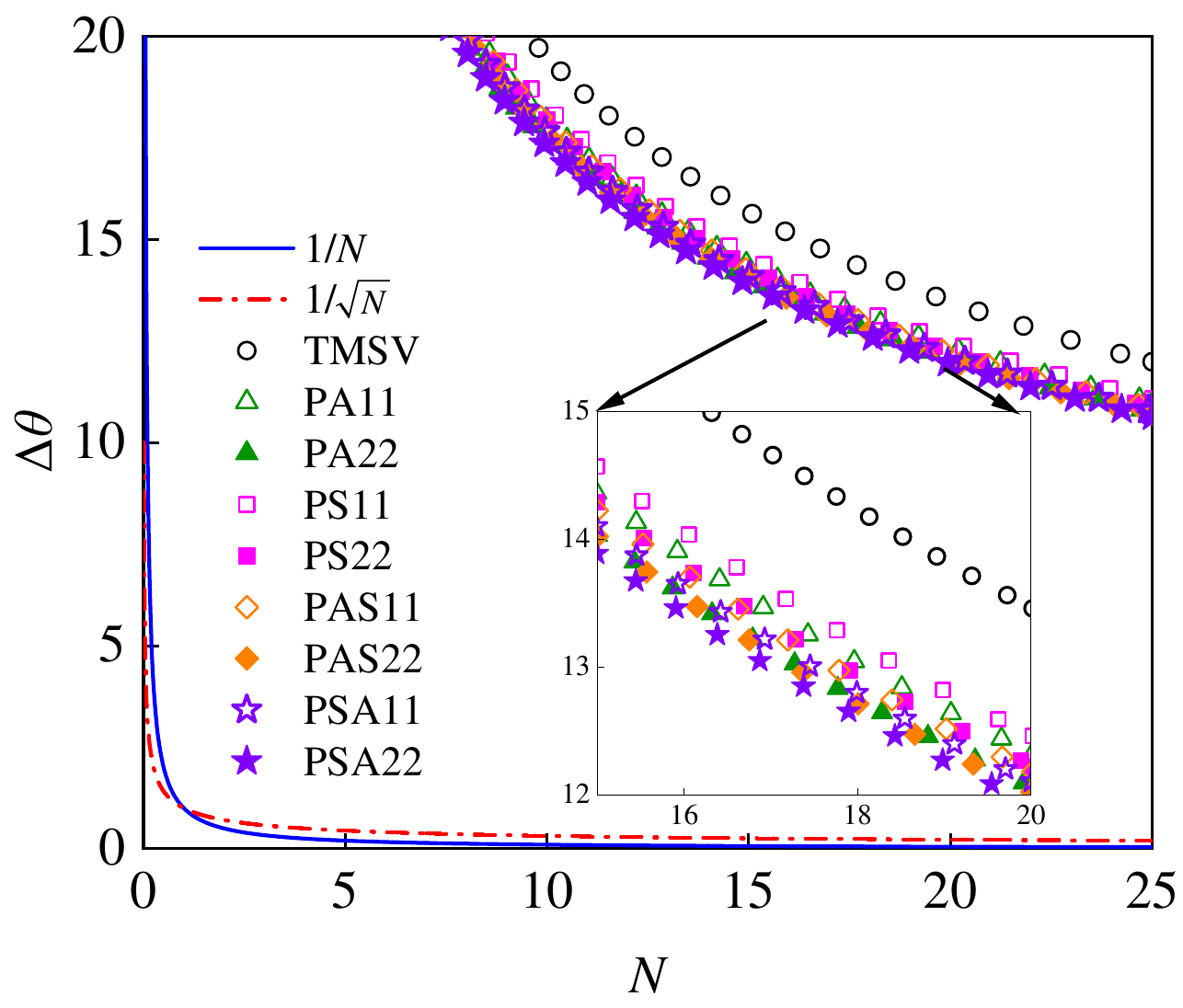}
		\caption{\textcolor{blue}{The phase estimation sensitivity $\Delta \theta$ versus mean photon number $N$ for all input states. The transmittance parameter has been set to $T=0.5$ for all cases. The phase shift value has been taken as $\theta = 10^{-4}$. } }
		\label{f9}
	\end{figure}
	We also show $\Delta \theta$ against $N$ with the influence of the photon loss, as depicted in Fig.~\ref{f9}. Both Gaussian and non-Gaussian states own sensitivity far from $1/N$ and $1/\sqrt{N}$ limits. Another result of this figure is that the TMSV state is fragile to the influence of photon loss, and the PSA state shows its robustness. With $T=0.5$, the TMSV state shows the worst $\Delta \theta$, though it is the best input state without the noise, according to the Fig.~\ref{f2}, Fig.~\ref{f3}, and Fig.~\ref{f5}.
	
	\subsection{OAM mitigates the influence of the photon loss}\label{S3C}
	
	From the discussion above, our results demonstrate that the increase of noise causes deterioration to the phase estimate for small phase shifts. To mitigate the deterioration brought by photon loss, we employ non-Gaussian state with the enhancement from OAM.
	
	\begin{figure}[htb]
		\centering
		\includegraphics[scale=0.39]{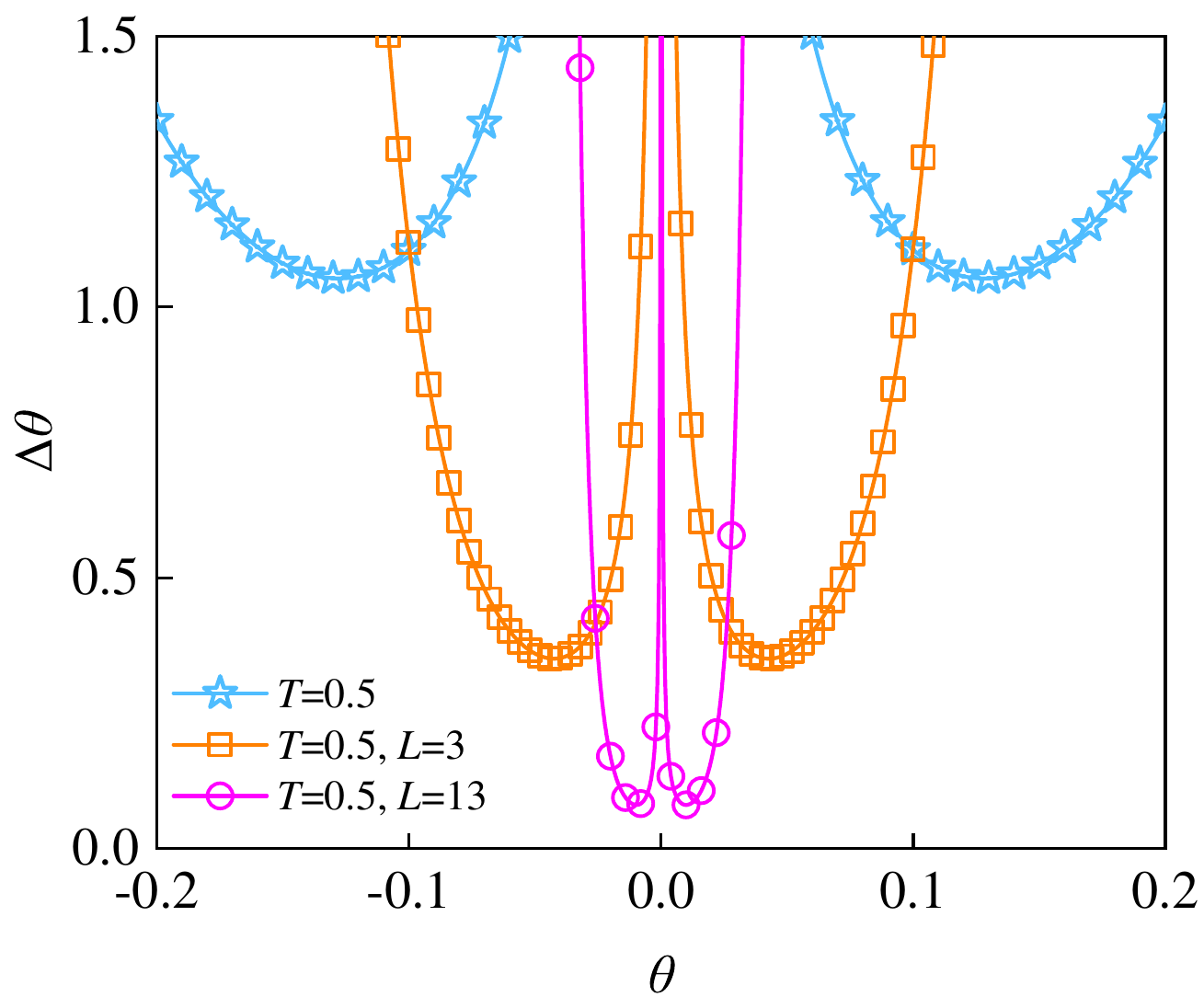}
		\caption{ \textcolor{blue}{The phase estimate sensitivity $\Delta \theta$ as a function of phase $\theta$ PSA11 state. The squeezing parameter has been taken as $\emph r=1.096$. The star-blocked solid line represents the measurement with $T=0.5$ without enhancement from OAM. The square-blocked and circle-blocked solid denote the $\Delta \theta$ with transmittance parameter $T=0.5$ for OAM quantum numbers $L=3$ and $13$, respectively. }}
		\label{f10}
	\end{figure}
	
	A possible explanation for the ``gap" with the presence of photon loss might be that the noise could be magnified at the near decorrelation point $(\theta \rightarrow0)$~\cite{zhang2021improving,li2018effects,Chen16}. As shown in Fig.~\ref{f10}, with the increase of topological charge $L$, the sensitivity $\Delta \theta$ reaches lower bounds for transmittance $T=0.5$. At the same time, the ``gap'' of the curve for $\theta \rightarrow 0$ is getting narrower with the enhancement of $L$, which makes it easy to measure small phase shifts even with $50\%$ photon loss in both arms of the MZI. The figure reveals the fact that we can further mitigate the deterioration with OAM.

	\begin{figure}[htb]
		\centering
		\includegraphics[scale=0.35]{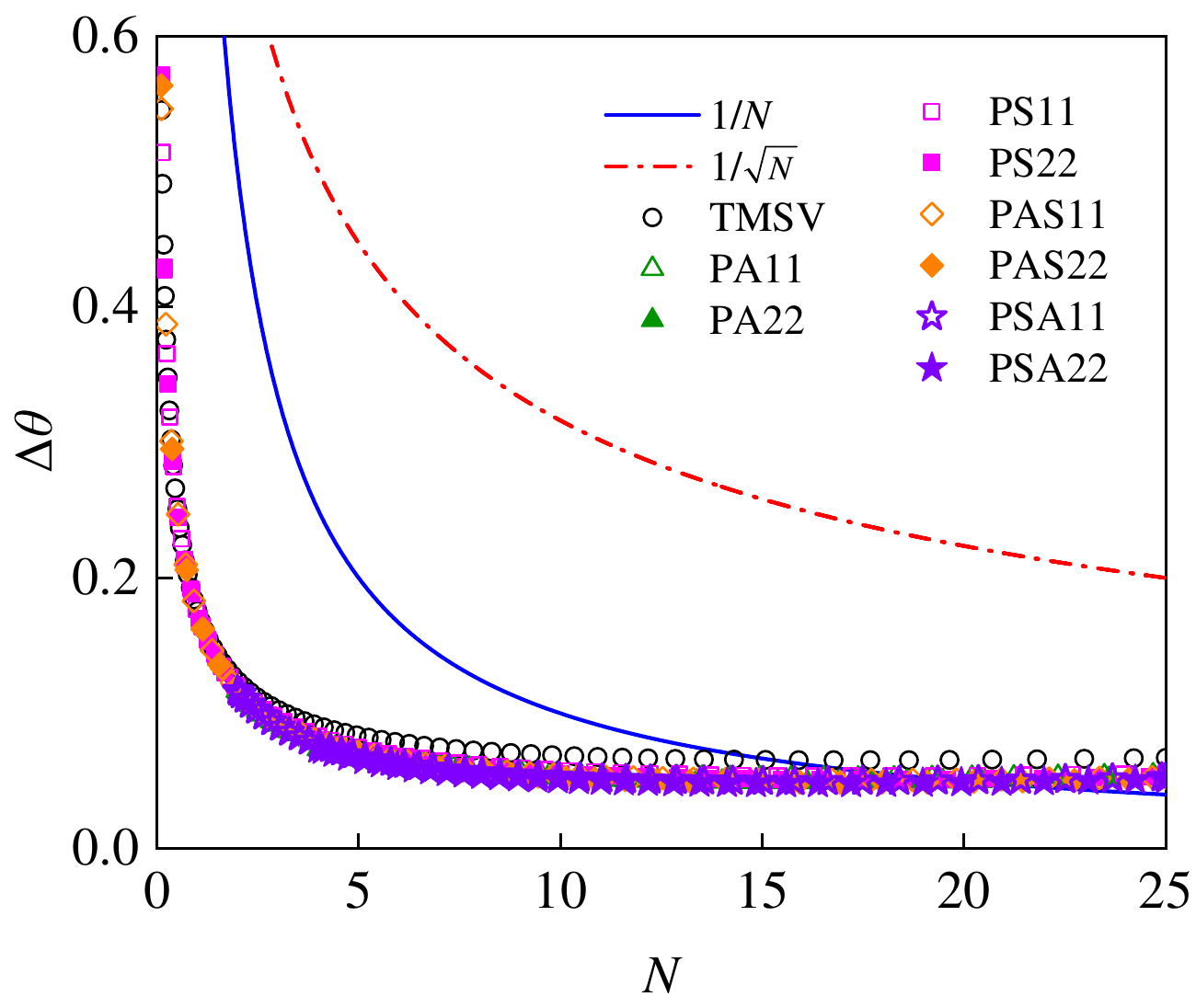}
		\caption{\textcolor{blue}{Phase estimation sensitivity of entanglement states $\Delta \theta$ as a function of mean photon number $N$ for all input states. The phase shift $\theta$ and OAM quantum number $L$ have been taken as $\theta=0.007$ and $L=21$, respectively. The solid blue and red dash-doted lines represent HL and SNL, respectively.  }}
		\label{f11}
	\end{figure}
	
	Significant enhancement of OAM contributed by SPPs~\cite{wang2016study} makes it impossible to estimate small phase shifts with noise. Fig.~\ref{f11} demonstrates the sensitivity $\Delta \theta $ versus mean photon number $N$ with the $50\%$ photon loss for $L=21$. With the enhancement of OAM, the $\Delta \theta$ surpasses the HL when $N \textless 15$ for all input state candidates. Remarkably, the enhancement of OAM can mitigate the deterioration brought by photon loss.
	
	\begin{figure}[htb]
		\centering
		\includegraphics[scale=0.35]{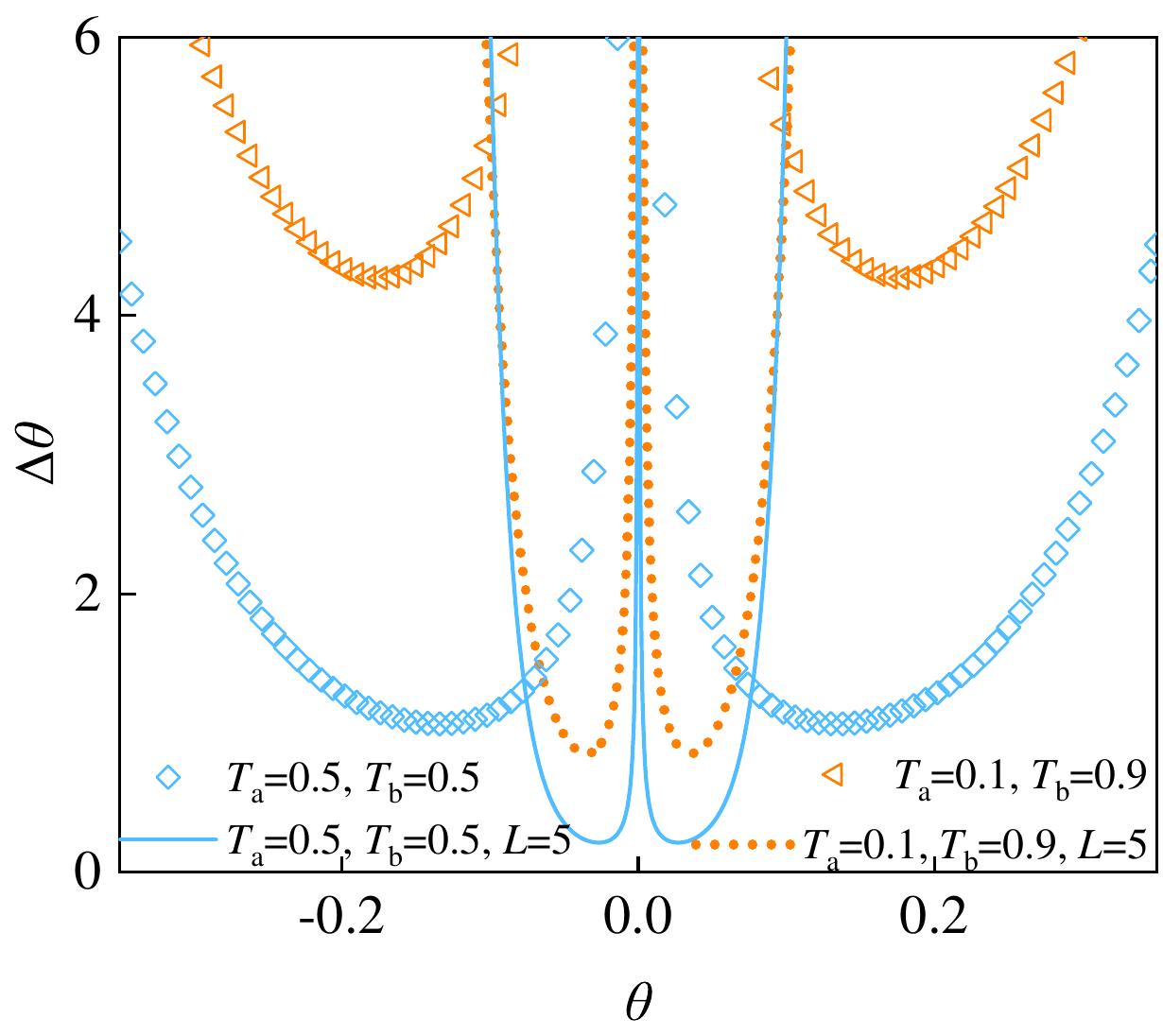}
		\caption{\textcolor{blue}{Phase estimation sensitivity $\Delta \theta$ as a function of phase shift $\theta$ PAS11 state, the squeezing parameter has been taken as $r=1.096$. The diamond icons and solid line denote the measurement without the enhancement of OAM and $L=5$, and the transmittance parameter is set to $T=0.5$. Triangle and dots represent the phase shifts with non-enhancement from OAM and the OAM quantum number taken as $L=5$, where the asymmetric photon loss $T_{\rm a(\rm b)}=0.1(0.9)$.}}
		\label{f12}
	\end{figure}
	
	In order to contrast the impact of OAM enhancement on symmetric and asymmetric photon loss, we present Fig.~\ref{f12}. The findings reveal that, for a constant topological charge $L$, the phase estimation sensitivity performs more optimally with symmetric noise as compared to asymmetric noise.

	\begin{figure}[htb]
		\centering
		\includegraphics[scale=0.35]{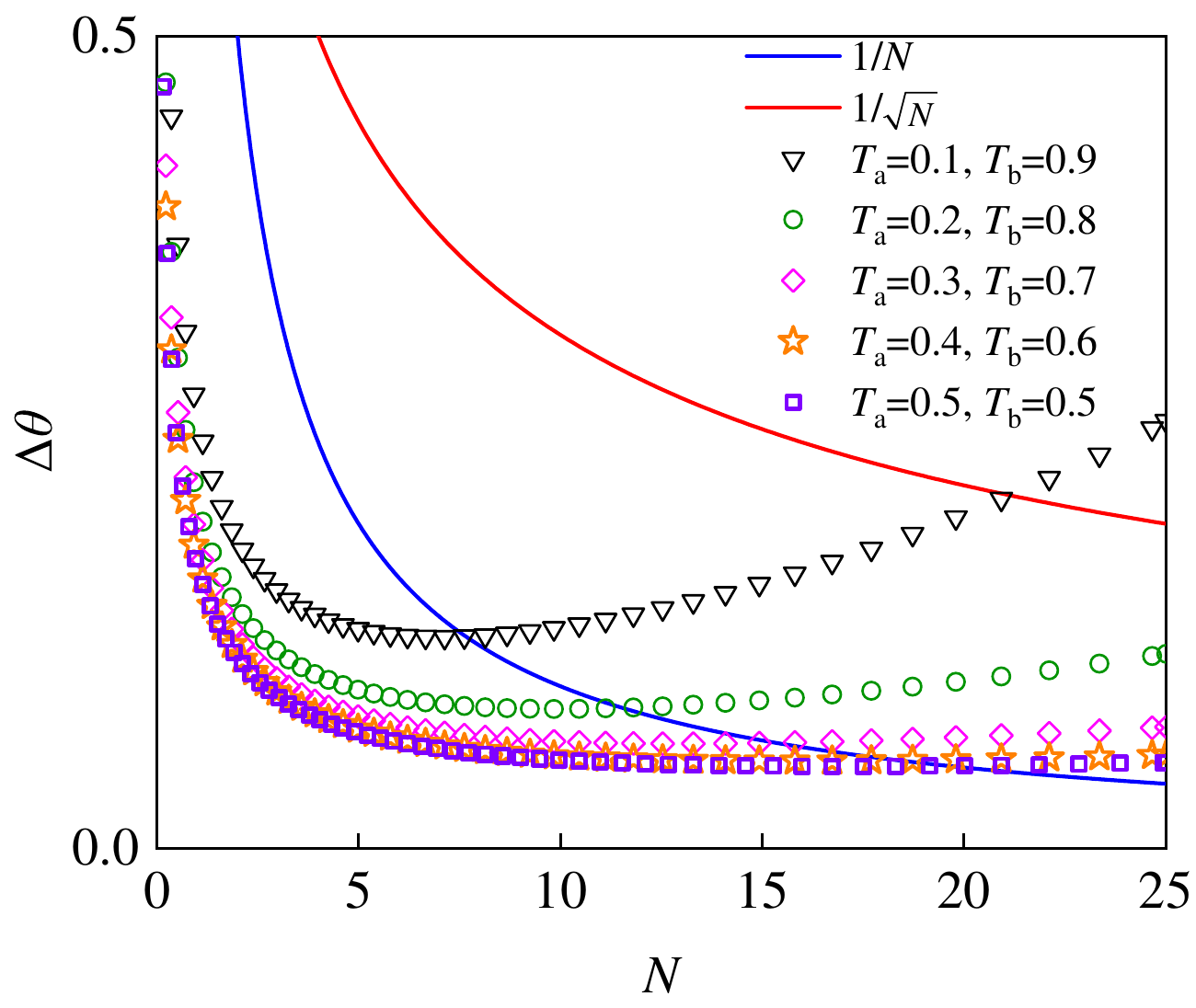}
		\caption{\textcolor{blue}{Phase estimation sensitivity $\Delta \theta$ versus mean photon number $N$ with various photon loss for PAS11 state. The phase shift value and OAM quantum number have been set as $\theta =0.007$ and $L=21$.  }}
		\label{f13}
	\end{figure}

	To display the effectiveness of OAM enhancement, we plot the sensitivity $\Delta \theta$ as the function of mean photon number $N$ with various photon loss. The result from this figure is clearly that for $\theta =0.007$, the symmetric photon loss owns the best sensitivity which reveals similar conclusion with Fig.~\ref{f7}. With the topological charge $L=21$, the phase estimation sensitivity can readily surpass the $1/N$ limit, without necessitating a significantly higher mean photon number.

	\section{\label{sec:level2}Statistical properties of Gaussian and non-Gaussian states}
	
\begin{figure}[htb]
		\centering
		\renewcommand{\thefigure}{14}
		\includegraphics[scale=0.39]{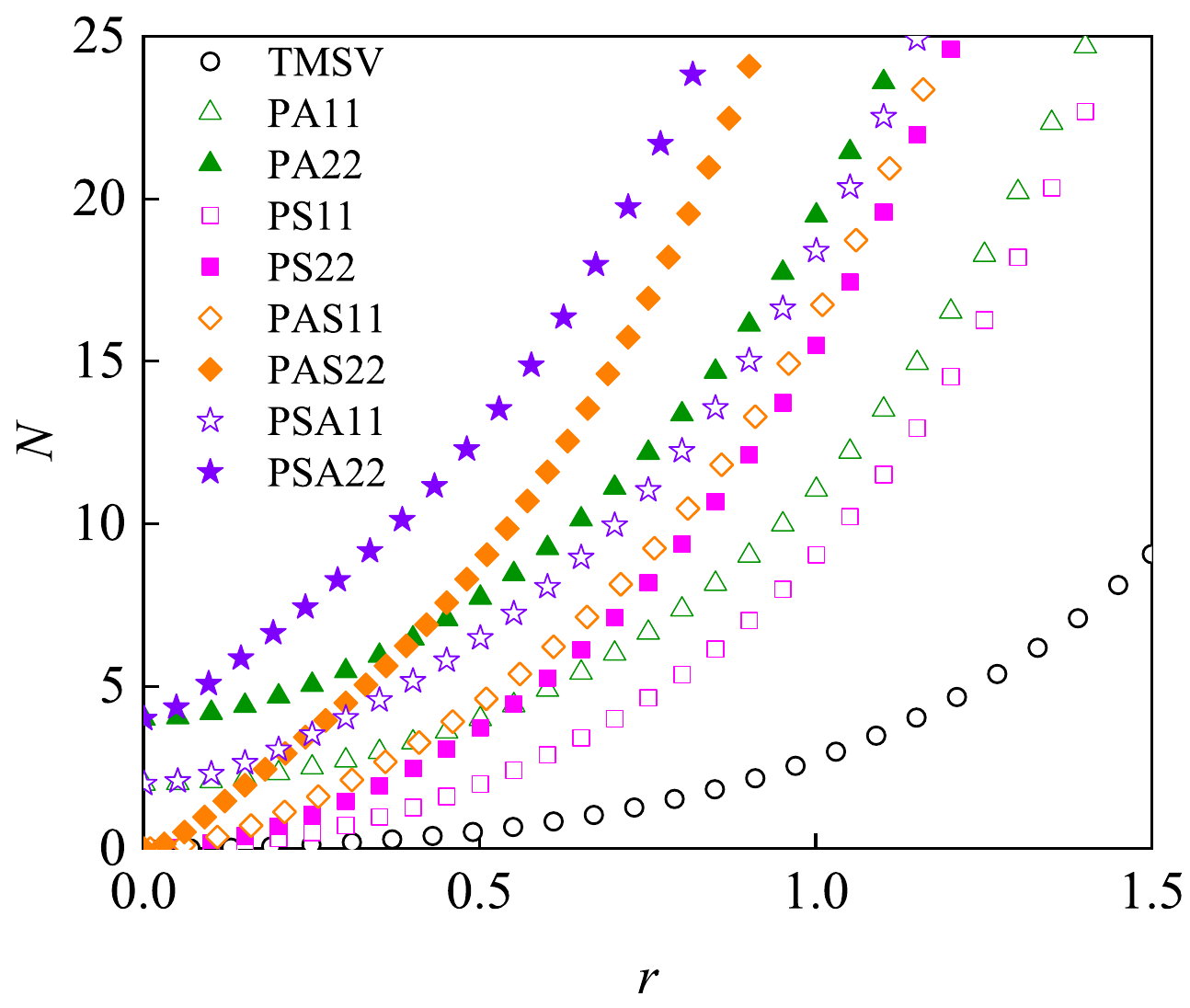}
		\caption{Mean photon number as a function of squeezing parameter for Gaussian and non-Gaussian states. Black circles denotes TMSV state; square, triangle, diamond and star indicate PS, PA, PAS and PSA, respectively. The hollow icons represent low order $G=H=1$ and solid icons for $G=H=2$. }
		\label{f14}
	\end{figure}

	In the protocol, the TMSV states can be described as~\cite{QI2018,PRXQuantum.2.030204} 
\begin{eqnarray}\label{e3}
	&|  \Psi \rangle_{\rm TMSV}&= \hat{S}(\xi) |n_{1},n_{2} \rangle \nonumber\\ &&=  \rm{exp}(\xi ^{*} \hat{a} \hat{b}-\xi \hat{a}^{\dagger} \hat{b}^{\dagger}  ) |n_{1},n_{2} \rangle 
\end{eqnarray}	
	 where $\hat{S}(\xi)$ represent two-mode squeezing parameter, $\xi=re^{i\psi}$, and r is know as the squeezing parameter and $\psi$ as squeezing angle. We illustrate the mean photon number of the proposed states as a function of the squeezing parameter in Fig.~\ref{f14}. As shown in the figure, when $r \rightarrow 0$, the initial $N$ of non-Gaussian state is determined by the Bose operator acting on the TMSV state. The plot further elucidates that for a given $r$, the TMSV state has the lowest $N$, while PSA22 has the highest $N$.

	The properties of input states determine the result of measurement. We use von Neumann entropy as a witness of non-classical and entanglement~\cite{PhysRevLett.122.210402, 2020entropy}. Eq.~(\ref{e11}) gives the equation of entropy, where $\rho_{a}=\rm Tr_{b}[|\Psi_{ab} \rangle \langle \Psi_{ab}   | ]$ and  $\rho_{b}=\rm Tr_{a}[|\Psi_{ab} \rangle \langle \Psi_{ab}   | ]$. We use the label $E$ to represent entropy hereafter,
	\begin{equation}\label{e11}
		E(| \Psi_{\rm ab}  \rangle) = -\rm Tr[\rho_{\rm a} {ln}(\rho_{\rm a})]=-\rm Tr[\rho_{\rm b}  {ln}(\rho_{\rm b})].
	\end{equation}
	
	We use Eq.~(\ref{e11}) to obtain the entropy as a function of $r$ and show the result in Fig.~\ref{f14}. We can see from this plot that TMSV has the lowest entropy~\cite{PhysRevA.59.1820} while PAS22 and PSA22 own the largest.

	\begin{figure}[htbp]
		\centering
		\renewcommand{\thefigure}{15}
		\includegraphics[scale=0.39]{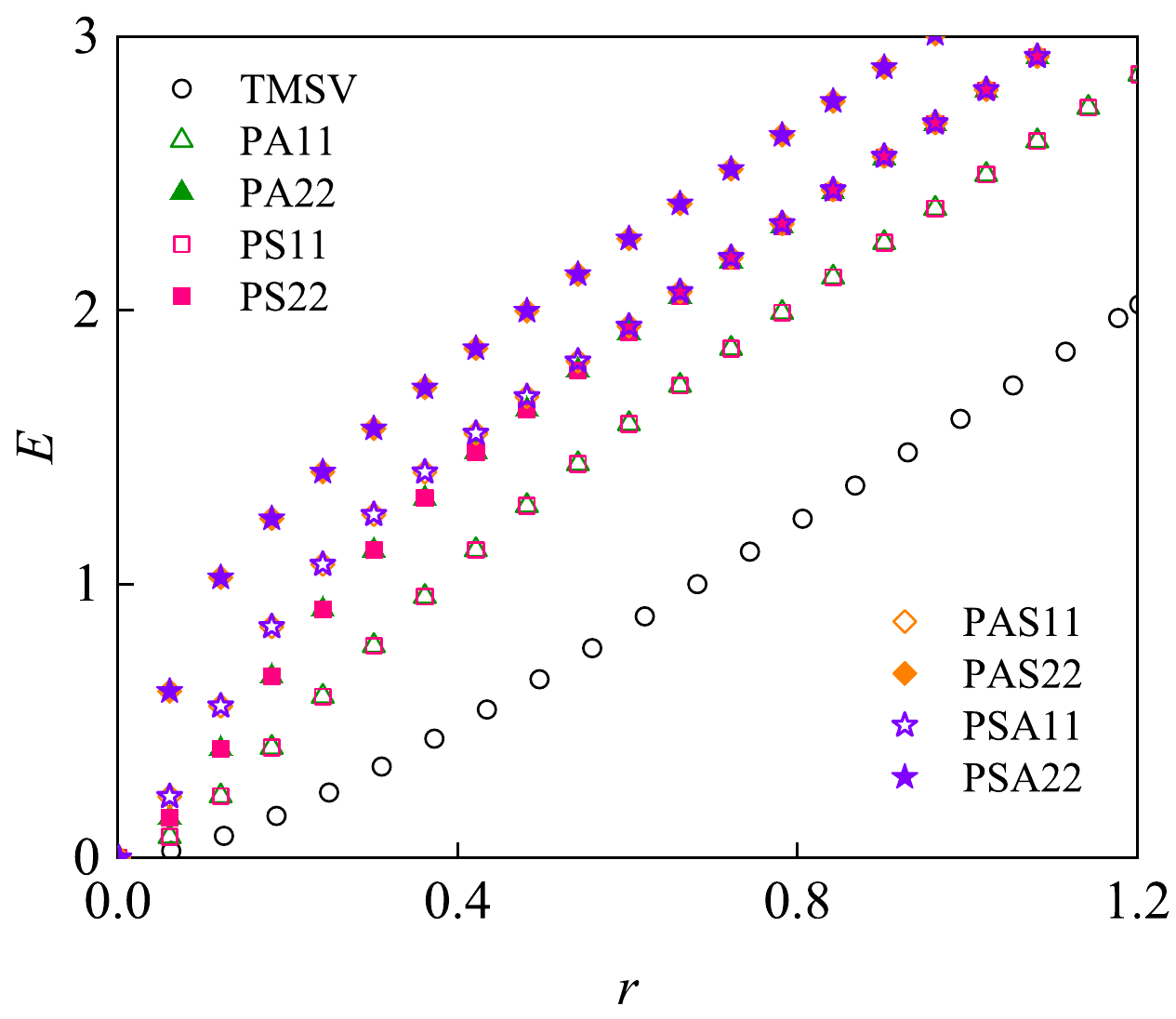}
		\caption{Von Neumann entropy as a function of mean photon number. The black circle denotes the TMSV state. PA, PS, PAS, and PSA are represented by triangle, square, diamond and star, respectively. The hollow icon represented low order $G=H=1$ and the solid icon for $G=H=2$.}
		\label{f15}
	\end{figure}

In this study, we employ parity measurement to enable high-sensitivity measurement. A parity measurement in one output mode is relevant to the detection of $ \sum\limits_{N=0}^{\infty} \sum\limits_{M=0}^{N} |N-M,M\rangle\langle M,N-M |$~\cite{2010PRL,2020zubairy_QKD}. Consequently, the properties of the input state can be analyzed via joint photon number distribution. In Fig.~\ref{f16}, our primary focus lies with the PAS and the PSA states due to their superior robustness in the presence of photon loss. Unlike the TMSV state, which exhibits a 'thermal-like down the diagonal' distribution~\cite{joint,Gerry2010,Carranza:12}, the PAS and the PSA states present a markedly different perspective. The maximum value of probability is located at $ |8,8\rangle$ (PAS) and $ |9,9\rangle$ (PSA). This distribution thus serves as a useful tool for distinguishing the non-Gaussian states we are working with.
	
  	\begin{figure*}[htbp]\centering
  	\renewcommand{\thefigure}{17}
  	\centering
  	\includegraphics[width=0.89\linewidth]{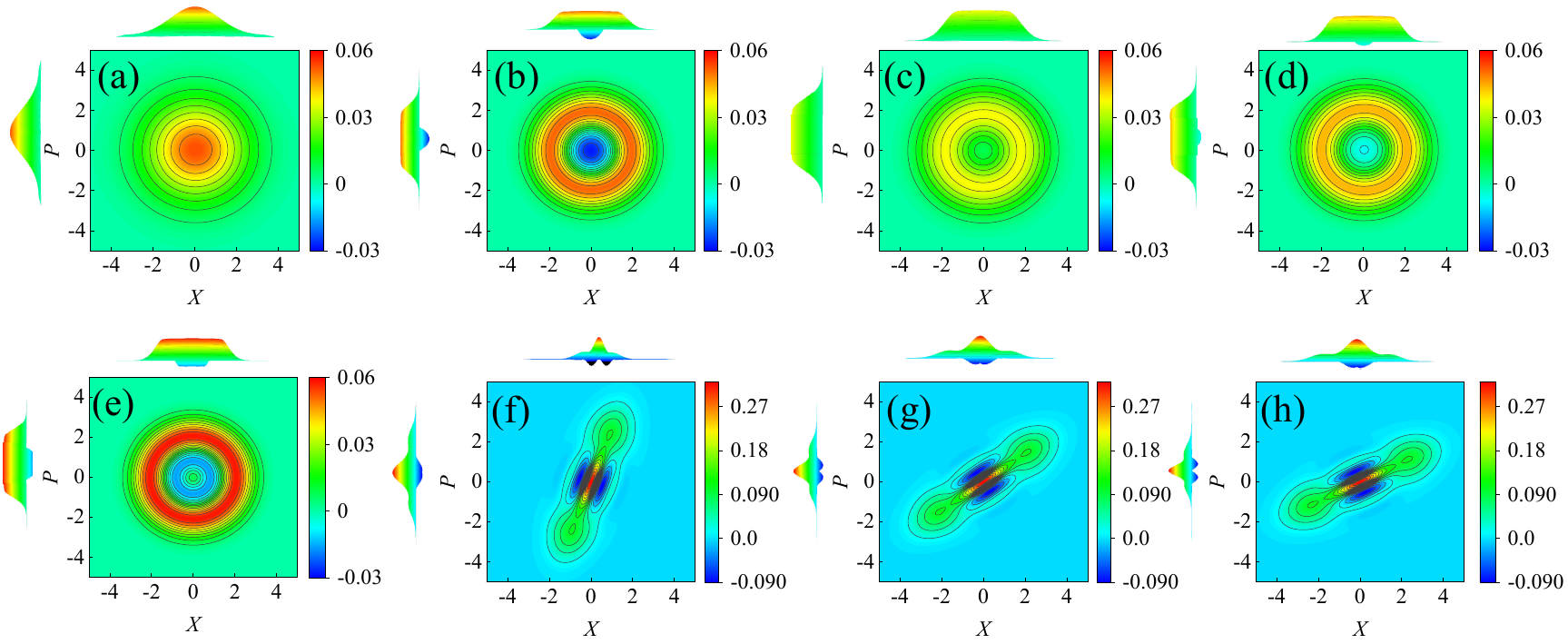}
  	\caption{Wigner function plot of input states with mean photon number $N=5$ and their evolution with different OAM. From panle (a) to panel (e) represent the wigner function plot of the TMSV, PA, PS, PAS, and PSA state. Data in panel (f) to panel (h) indicate the PSA state after phase shift with no OAM-enhancement, $L=7$ and $L=51$.In the top and left views of the plan diagram, we have provided the orthogonal and side views of the Wigner function, respectively.}
  	\label{f17}
  \end{figure*}
	
	\begin{figure}[htb]
		\centering
		\renewcommand{\thefigure}{16}
		\includegraphics[scale=0.35]{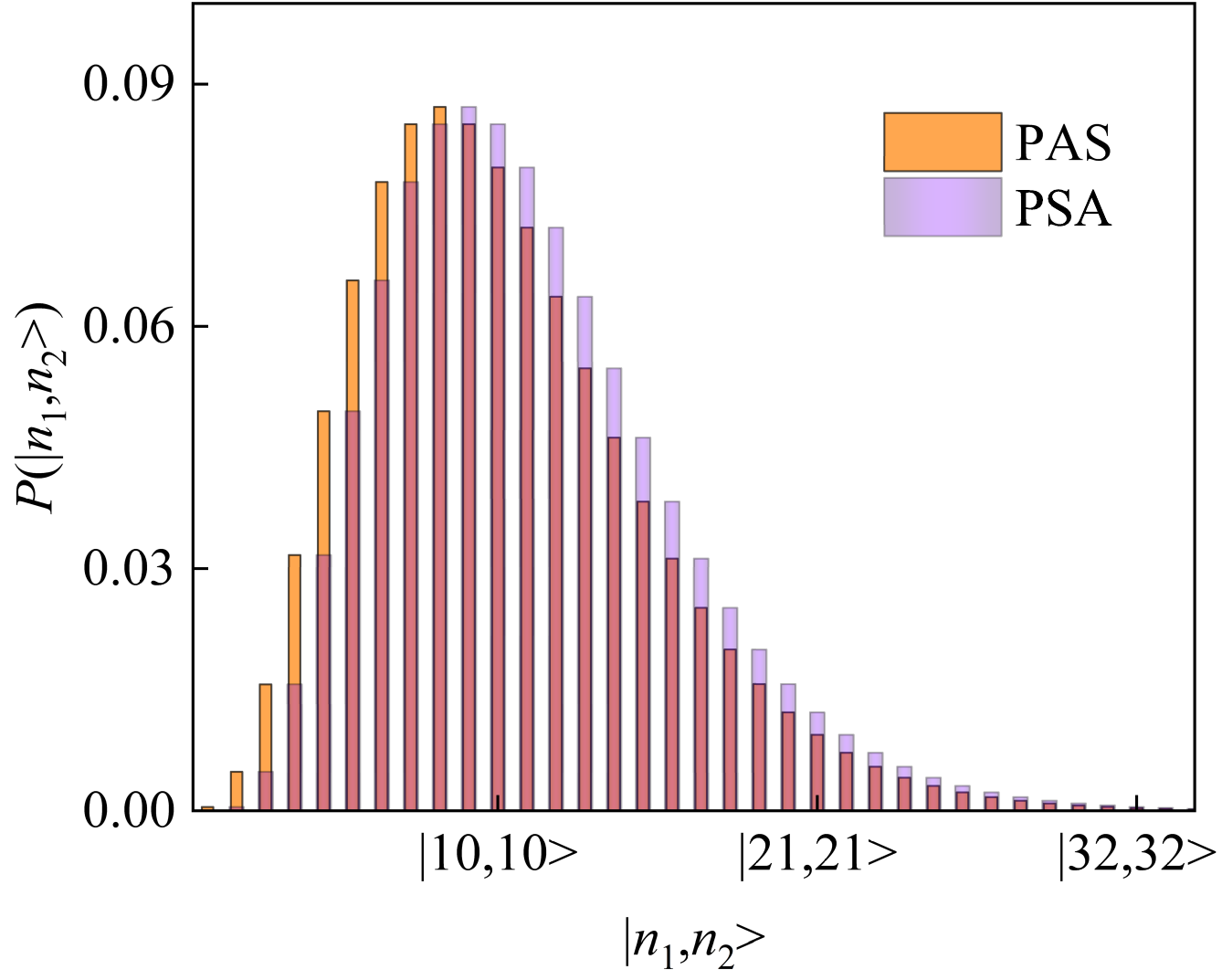}
		\caption{The joint photon number distribution for the PAS (ruby-colored) and PSA (blue-colored) state with the squeezing parameter is taken as $r =1.096$.}
		\label{f16}
	\end{figure}

To visualize the properties and distinguish different state and evolution of the input states, we use the Wigner function~\cite{zhang2021improved, qutip1, qutip2, newWIGNER1, newWIGNER2, WIGNER3,WIGNER4}. As Fig.~\ref{f17} shows, we compare the initial input state in the first row from panel (a) to (e). As we can see, non-Gaussian states owns negativity value of the wigner function in the center of the axis which exhibits their non-classical properties. Among all input states, the PSA state owns largest area of negativity which indicates it has more non-classical properties than the other states. This plot offers a useful way to identify the state owns the best sensitivity with the presence of photon loss. In the second row from panel (f) to (h), we compare the wigner function of the PSA state with different OAM number after the first BS. As we can see from figure, data in panel (f) indicates the state with no enhancement from OAM, panel (g) and (h) present state with OAM of $L=7$ and $L=51$. With higher OAM, the plot of state rotate with a larger angle.

	\section{Conclusion}\label{S7}

In summary, we utilize OAM and non-Gaussian states to achieve higher sensitivity in phase estimation under significant noise. The non-Gaussian state, particularly the PSA state exhibits higher sensitivity than the TMSV state. In order to further improve the sensitivity, we can balance the photon loss in the two arms of MZI. With the presence of the symmetric noise, non-Gaussian states with higher-order Bose operators achieve a lower sensitivity bound for small phase shift in the symmetric photon loss.

As noise levels escalate, the task of estimating small phase shifts becomes progressively challenging. Nevertheless, an increase in topological charge results in the photon carrying a higher degree of OAM. This enhancement improves the resilience of the estimation scheme by mitigating the detrimental effects of noise, thereby facilitating the estimation of small phase shifts ($\theta \rightarrow 0$). This makes it feasible to achieve the $1/N$ limit, even in the face of substantial photon loss, such as $50\%$ photon loss. Our research offers practical methodologies for realizing superior sensitivity in phase estimation under significant noise conditions. We anticipate that our findings will serve as a valuable and effective instrument in the field of quantum metrology.
	
	\section*{ACKNOWLEDGEMENTS}
	This work was supported by National Natural Science Foundation of China (NSFC) (Grant No.11675046), Program for Innovation Research of Science in
	Harbin Institute of Technology (Grant No. A201412), and Postdoctoral Scientific Research Developmental Fund of Heilongjiang Province (Grant No.LBH-Q15060).
	\\
	\appendix
	\section{sensitivity by parity detection and the QCRB}\label{a1}
	
	As high sensitivity quantum metrology, the Quantum Fisher Information (QFI) determined well-known quantum Cramér-Rao bound (QCRB) has been studied in phase estimate. The QCRB can be obtained according to input resource, and it is independent of detection, $\Delta \theta_{\rm QCRB} =1 / \sqrt{F_{\rm Q}}$, where $F_{\rm Q}$ denotes QFI. According to this theory, greater $F_{\rm Q}$ can achieve higher $\Delta \theta$. For an input state, the $F_{\rm Q}$ can be calculated as
	\begin{equation}\label{eA2-1}
		F_{\rm Q}=4\left[ \langle \hat{{\Psi}}^{'}\mid \hat{{\Psi}}^{'} \rangle  - \langle   \hat{{\Psi}}^{'}\mid \hat{{\Psi}}   \rangle^{2}    \right],
	\end{equation}
	where $\left| \Psi \right> = U_{\theta} U_{\rm BS1} | \Psi_{in} \rangle $ represent the state vector after the $\rm BS_{1}$ and the DP and prior to the second beam splitter $\rm BS_{2}$, and $| \hat{\Psi}^{'} \rangle = \partial \left| \Psi \right> / \partial \theta    $  Eq.~(\ref{eA2-1}) becomes
	\begin{equation}\label{eA2-2}
		F_{Q}=4\left[ \left< \Psi_{\rm in}\mid \hat{J}_{2}^2 \mid \Psi_{\rm in} \right>  - |\left< \Psi_{\rm in}\mid \hat{J}_{2} \mid \Psi_{\rm in} \right>  |^{2}   \right],
	\end{equation}
	where $J_{1,2,3}$ represent the angular momentum operators shows in the Eq.~(\ref{eA2-3}),
	\begin{equation}\label{eA2-3}
		\begin{split}
			\begin{aligned}
				{\hat{J}}_{1}&= {\frac{1}{2}} {\left(\hat{a}^{\dagger} \hat{b}+\hat{a} \hat{b}^{\dagger}\right) },{\hat{J}}_{2}= {\frac{1}{2i}} {\left(\hat{a}^{\dagger} \hat{b}-\hat{a} \hat{b}^{\dagger}\right) },\\{\hat{J}}_{3}&= {\frac{1}{2}} {\left(\hat{a}^{\dagger} \hat{a}- \hat{b}^{\dagger} \hat{b}  \right) }.
			\end{aligned}
		\end{split}
	\end{equation}
	
\begin{figure}[!h]
		\centering
		\includegraphics[scale=0.32]{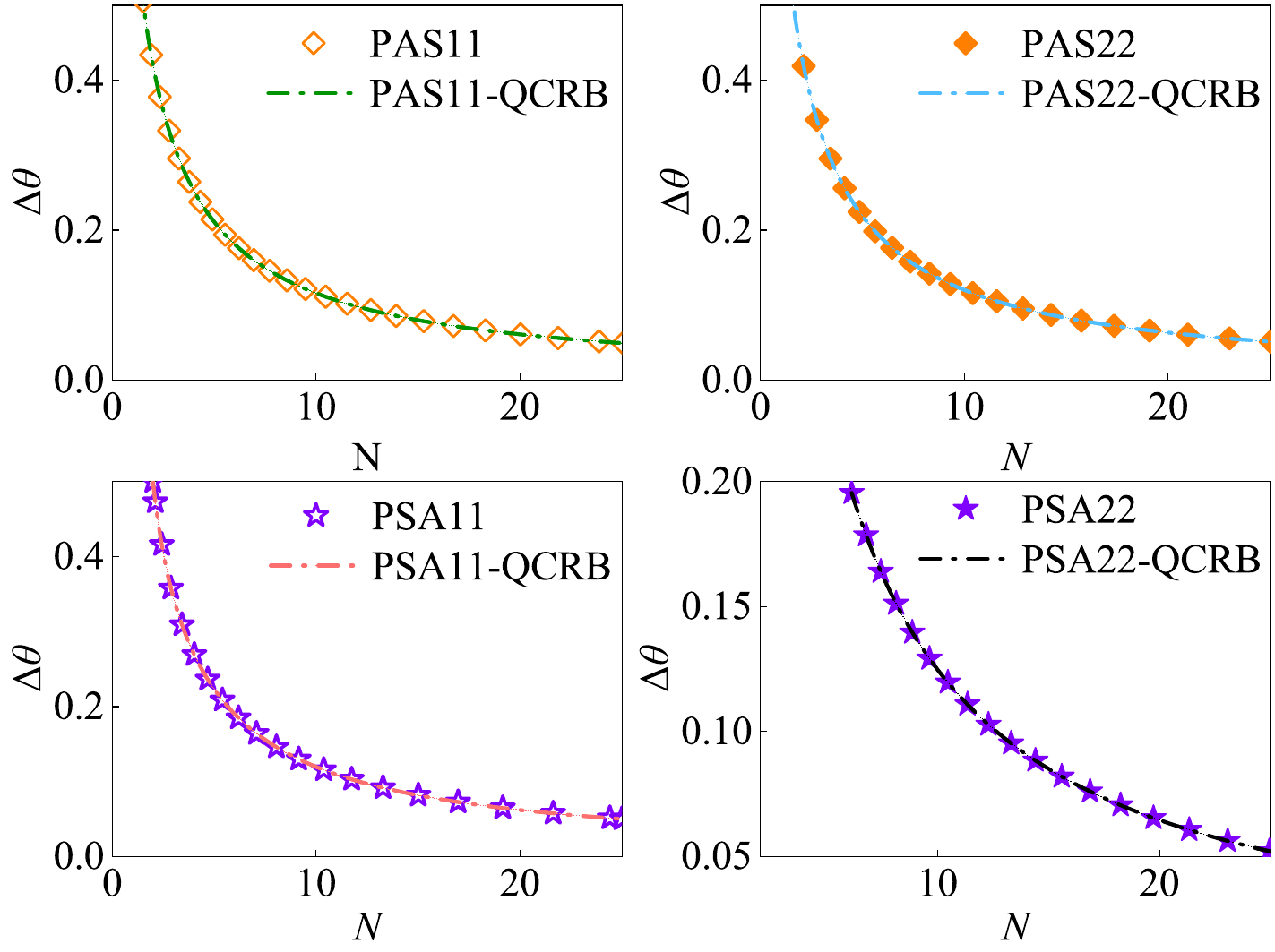}
		\caption{Phase estimation sensitivity of the PAS and PSA states and their QCRB in the absence of noise. Diamonds and stars represent the PAS state and the PSA state. The hollow icons represent low order $G=H=1$ and solid icons for $G=H=2$.}
		\label{fA1}
	\end{figure}

	To thoroughly investigate the potential of the input resource, we plot the QCRB and sensitivity versus the mean photon number $N$ in Fig.~\ref{fA1}. We use diamonds and stars to represent the sensitivity of different input states and a dash-doted line to represent the QCRB of the corresponding state. The hollow icons denote the state with $G=H=1$, and solid icons represent states with $G=H=2$. These results suggest that parity offers a good detection that can saturate the QCRB.
	
	\section{Derivation of parity detect signal and its sensitivity }\label{a2}
	
	The non-Gaussian state is obtained by harnessing the Bose operator on the TMSV state represented by
	\begin{eqnarray}\label{eA1}
		&{|\hat{\Psi}\rangle}_{\rm PAS}=& {\hat{a}^{\dagger G}}{\hat{b}^{H}}{\hat{S}}{\left|0,0\right>},\nonumber \\ 
		&{|\hat{\Psi}\rangle}_{\rm PSA}= &{\hat{a}^{ G}}{\hat{b}^{\dagger H}}{\hat{S}}{\left|0,0\right>}.
	\end{eqnarray}
	
	In the protocol, we use the parity detection method. The parity operator can be represented in coherence state, as shown in Eq.~(\ref{eA2}), where $\left| \gamma \right> $ is the coherent state.
	\begin{equation}\label{eA2}
		\hat{\Pi}_{b} = (-1)^{\hat{n}} = e^{i\pi {\hat{b}^\dagger} \hat{b}} = \int \frac{\,d^2 x}{\pi} \left| \gamma \right> \left< -\gamma \right|.
	\end{equation}
	
	After harnessing the parity operator on the output state, the expectation value of detection signal can be represented as
	\begin{equation}\label{eA3}
		\left< \Pi_{b} \right> = {}^{}_{\rm out}\! \left<\Psi \right|  \int \frac{\,d^2 x}{\pi} \left| \gamma \right> \left< -\gamma \right|  \left| \Psi \right>_{\rm out}.
	\end{equation}
	
	We can also represent gaussian and non-gaussian states in coherence state, here we take the TMSV as an example, where $z=\rm tanh(\emph r)$
	
	\begin{eqnarray}\label{eA4}
		&\left|  \Psi\right>_{\rm{TMSV}} =& (1-z^{2})^{\frac{1}{2}}  \int \frac{\,d^2 \alpha \,d^2 \beta }{\pi^{2}} \rm exp \Big[ - \frac{|\alpha|^{2} |\beta|^{2}  }{2}\nonumber \\&& +  {\alpha}^{*}{\beta}^{*}z  \Big]  \times \left| \alpha , \beta  \right> .   
	\end{eqnarray}	
	
The evolution can be represented by a unitary operator, the evolution of the input state can be simplified as the evolution of Bose operator. After representing input states in the basis of the coherent state, the state's evolution can be calculated as follows
	\begin{equation}\label{eA5}
		e^{\xi \hat{A}} \hat{B} e^{- \xi \hat{A}} = \hat{B}+\xi [\hat{A},\hat{B}]+\frac{{\xi}^{2}}{2!}[\hat{A},[\hat{A},\hat{B}]]+\ldots \ .
	\end{equation}
	
	 With the absence of the photon loss, the evolution can be represented as 
	\begin{eqnarray}\label{eA6}
		U^{\dagger}_{\rm{MZI}} \hat{a}^{\dagger} U_{\rm{MZI}} = \hat{a}^{\dagger}\mathrm{cos\frac{\theta}{2}}  + \hat{b}^{\dagger} \mathrm{sin\frac{\theta}{2}},\nonumber \\
		U^{\dagger}_{\rm{MZI}} \hat{b}^{\dagger} U_{\rm{MZI}} = \hat{b}^{\dagger}\mathrm{cos\frac{\theta}{2}}  - \hat{a}^{\dagger} \mathrm{sin\frac{\theta}{2}}.
	\end{eqnarray}	
	
In order to calculate photon loss, we introduce two fictitious Beam Splitters (BSL1 and BSL2) into the path of the MZI system. After the evolution, including photon loss, we obtain the reduced density matrix by tracing over the environmental modes that have passed through the beam splitter. In the presence of the photon loss, the evolution of lossy MZI protocol can be represented as~\cite{PRXQuantum.3.010202}
	\begin{eqnarray}\label{eA7}
		&U^{\dagger}_{\rm MZI} \hat{a}^{\dagger} U_{\rm MZI} = &  \frac{1}{2} [ \hat{a}^{\dagger} \mathrm{cos\frac{\theta}{2}} \mathrm{(cosQ_{\rm a}+cosQ_{\rm b} )} \nonumber \\&& +i \hat{b}^{\dagger} \mathrm{cos\frac{\theta}{2}} \mathrm{ (cosQ_{\rm a}-cosQ_{\rm b})} \nonumber \\&&+i\hat{a}^{\dagger} \mathrm{sin\frac{\theta}{2}} \mathrm{(cosQ_{\rm b}-cosQ_{\rm a})} \nonumber \\&&+\hat{b}^{\dagger}\mathrm{sin\frac{\theta}{2}} \mathrm{(cosQ_{\rm b}+cosQ_{\rm a}) }        ]        ,\nonumber \\
		&U^{\dagger}_{\rm MZI} \hat{b}^{\dagger} U_{\rm MZI} =& \frac{1}{2} [i \hat{a}^{\dagger} \mathrm{cos\frac{\theta}{2}} \mathrm{ (cosQ_{\rm b}-cosQ_{\rm a} )} \nonumber \\&& + \hat{b}^{\dagger} \mathrm{cos\frac{\theta}{2}}\mathrm{ (cosQ_{\rm a}+cosQ_{\rm b})} \nonumber \\&&+\hat{a}^{\dagger} \mathrm{sin\frac{\theta}{2}} \mathrm{ (-cosQ_{\rm a}-cosQ_{\rm b})} \nonumber \\&&+ i \hat{b}^{\dagger}\mathrm{sin\frac{\theta}{2}}\mathrm{(cosQ_{\rm b}-cosQ_{\rm a}) }        ] ,
	\end{eqnarray}	
	where $\mathrm{cos^{2}Q_{\rm a}}= T_{\rm a}, \mathrm{ cos^{2}Q_{\rm b}}= T_{\rm b}$, By using the Eq.~(\ref{eA7}) and integral below, we can derive the parity detection signal in Eq.~(\ref{eA8}),
	
	\begin{eqnarray}\label{eA8}
		\int \frac{\,d^2 z }{\pi} {\rm exp}[\zeta |z|^{2} +\xi z+{\eta}z^{*}+fz^{2}+gz^{*2} ] \nonumber \\ =\frac{1}{\sqrt{\zeta^{2}-4fg}}  {\rm exp}\left[ \frac{-\zeta \xi \eta+ \xi^{2}g + \eta^{2}f}{\zeta^{2}-4fg}       \right].
	\end{eqnarray}	
	
	According to Eq.~(\ref{eA3}), Eq.~(\ref{eA7}), and Eq.~(\ref{eA8}), we obtain the parity signal of the output state enhanced by OAM in a lossy MZI system. With the presence of noise, the equation will be too long to demonstrate. We assume $T=1$ (no noise) and give the equation of PSA11 sensitivity as
	
	\begin{eqnarray}\label{eA9}
		&\Delta\theta_{\rm PSA11(\theta + \pi /2)}= & \Big\{ 1-\{z^{4}(-1+z^{2})^{2}[-4*(2\nonumber\\&&-63z^{4}+39z^{8}+14z^{12})\rm cos\emph L(\pi\nonumber\\&&+2\theta)+\emph{z}^{2}(-4+315z^{4}-252z^{8}\nonumber\\&&+8z^{12}+4*(15-41z^{4}+9z^{8})\nonumber\\&& \rm cos2\emph L(\pi+2\theta)+4\emph{z}^{2}(-9+z^{4})\nonumber\\&&\rm cos3\emph L(\pi+2\theta)+\emph{z}^{4}\rm cos4\emph L(\pi\nonumber\\&&+2\theta)   )]^{2}       \} /  [64(1+z^{4}+2z^{2}\nonumber\\&&\rm cos2\emph L\theta)(1+z^{4}+2z^{2}cos\emph L(\pi\nonumber\\&&+2\theta))]^{8}  \Big\}^{\frac{1}{2}}   \Big/  \Big\{  2 \Big\vert \{ Lz^{2}(-1+z^{2})\nonumber\\&&[1-51z^{4}+396z^{8}-245z^{12}\nonumber\\&&+15z^{16}-6z^{2}(6-49z^{4}+10z^{8}\nonumber\\&&+10z^{12})\rm cos\emph L(\pi+2\theta)+3z^{4}(19\nonumber\\&&-20z^{4}+5z^{8})\rm cos2\emph L(\pi+2\theta)\nonumber\\&&-10z^{6}\rm cos3\emph L(\pi+2\theta)]\rm sin\emph L(\pi\nonumber\\&&+2\theta)     \} /   \{        (1+z^{4}+2z^{2}\nonumber\\&&\rm cos2\emph L\theta)^{\frac{1}{2}}[1+z^{4}+2z^{2}\mathrm{cos}\emph{L}(\pi\nonumber\\&&+2\theta)]^{5} \} \Big\vert \Big\}.
	\end{eqnarray}	
	
	The parity signal of PAS11 state is shown as 
	\begin{eqnarray}\label{eA10}
		&\left<  \Pi \right>_{\rm{PAS} (\theta + \pi /2)}=&\{ (1-z^{2})[8-252z^{4}+315z^{8}\nonumber\\&&-4z^{12}-4z^{2}(14+39z^{4}63z^{8}\nonumber\\&&+2z^{12})\rm cos\emph L(\pi+2\theta)+4\emph z^{4}(9\nonumber\\&&-41z^{4}+15z^{8})\rm cos2\emph L(\pi+2\theta)\nonumber\\&&+4\emph z^{6}\rm cos3\emph L(\pi+2\theta)-36z^{10}\rm \nonumber\\&&cos3\emph L(\pi+2\theta)+z^{8}\rm cos4\emph L(\pi\nonumber\\&&+2\theta)]       \} /  \{ 8(1+z^{4}+2z^{2}\nonumber\\&&\rm cos2\emph L\theta)^{\frac{1}{2}} (1+z^{4}\nonumber\\&&+2z^{2}\rm cos\emph L(\pi+2\theta))      \}  .
	\end{eqnarray}	
	
	The sensitivity of PAS11 state is represented as 
	\begin{eqnarray}\label{eA11}
		&\Delta\theta_{\rm PAS11 (\theta + \pi /2)}=& \Big\{ 1-\{(-1+z^{2})^{2}[8-252z^{4}\nonumber\\&&+315z^{8}-4z^{12}-4z^{2}(14+39z^{4}\nonumber\\&&-63z^{8}+2z^{12})\rm cos\emph L(\pi+2\theta)\nonumber\\&&+4z^{4}(9-41z^{4}+15z^{8})\rm cos2\emph L(\pi\nonumber\\&&+2\theta)+4z^{6}\rm cos3\emph L(\pi+2\theta)\nonumber\\&&-36z^{10}\rm cos3\emph L(\pi+2\theta)+z^{8}\nonumber\\&&\rm cos4\emph L(\pi+2\theta)       \} /  [64(1+z^{4}\nonumber\\&&+2z^{2}cos2L\theta)(1+z^{4}+2z^{2}\nonumber\\&&\rm cos\emph L(\pi+2\theta))]^{8}  \Big\}^{\frac{1}{2}}   \Big/  \Big\{  2 \Big\vert \{ \emph{L} \emph{z}^{2}\nonumber\\&&(-1+z^{2})[15-245z^{4}+396z^{8}\nonumber\\&&-51z^{12}+z^{16}-6z^{2}(10+10z^{4}\nonumber\\&&-49z^{8}+6z^{12})\rm cos\emph L(\pi+2\theta)\nonumber\\&&+3z^{4}(5-20z^{4}+19z^{8})\nonumber\\&&\rm cos2\emph L(\pi+2\theta)-10z^{10}\rm cos3\emph L(\pi\nonumber\\&&+2\theta)]\rm sin\emph L(\pi+2\theta)     \} /   \{        (1+z^{4}\nonumber\\&&+2z^{2}\rm cos2\emph L\theta)^{\frac{1}{2}}[1+z^{4}\nonumber\\&&+2z^{2}\rm cos\emph L(\pi2\theta)]^{5}      \}  \Big\vert   \Big\} .
	\end{eqnarray}

	With the presence of photon loss, e.g., $T_{a},T_{b}\neq 1$, the equation of PAS and PSA state is too long to present. We here present the sensitivity of the PS11 state as an example,

		\begin{eqnarray}\label{eA12}
			&\left< \Pi  \right>_{\rm PS11 (\theta + \pi /2)}=& \{ 8z^{2}(-1+z^{2})[-128+32z^{2}\nonumber\\&&-32T^{2}_{a}z^{2}+16T^{2}_{\rm a}z^{2}-32T^{2}_{\rm b}\nonumber\\&&+16T^{2}_{\rm b}z^{2}-12z^{4}+16T_{\rm a}z^{4}\nonumber\\&&+8T^{3}_{\rm a}z^{4}-2T^{4}_{\rm a}z^{4}-16\sqrt{T_{\rm a}T_{\rm b}}z^{4}\nonumber\\&&+64T^{3/2}_{\rm a}\sqrt{T_{b}}z^{4}+16T_{\rm b}z^{4}\nonumber\\&&+56T_{\rm a}T_{\rm b}z^{4}+8T^{2}_{\rm a}T_{\rm b}z^{4}\nonumber\\&&+64\sqrt{T_{\rm a}}T^{3/2}_{\rm b}z^{4}-16T^{3/2}_{\rm a}T^{3/2}_{\rm b}z^{4}\nonumber\\&&+8T_{\rm a}T^{2}_{\rm b}z^{4}-8T^{2}_{\rm a}T^{2}_{\rm b}z^{4}+8T^{3}_{\rm b}z^{4}\nonumber\\&&-2T^{4}_{\rm b}z^{4}-2T^{2}_{\rm a}z^{6}+2T^{3}_{\rm a}z^{6}-T^{4}_{\rm a}z^{6}\nonumber\\&&-8T^{3/2}_{\rm a}\sqrt{T_{\rm b}}z^{6}+8T^{5/2}_{\rm a}\sqrt{T_{\rm b}}z^{6}\nonumber\\&&-4T^{7/2}_{\rm a}\sqrt{T_{\rm b}}z^{6}-12T_{\rm a}T_{\rm b}z^{6}\nonumber\\&&+14T^{2}_{\rm a}T_{\rm b}z^{6}-6T^{3}_{\rm a}T_{b}z^{6}-8\sqrt{T_{\rm a}}\nonumber\\&&T^{3/2}_{\rm b}z^{6}+16T^{3/2}_{\rm a}T^{3/2}_{\rm b}z^{6}-4T^{5/2}_{\rm a}\nonumber\\&&T^{3/2}_{\rm b}z^{6}z^{6}-2T^{2}_{\rm b}z^{6}+14T_{\rm a}T^{2}_{\rm b} z^{6}\nonumber\\&&-2T^{2}_{\rm a}T^{2}_{\rm b}z^{6}+8\sqrt{T_{\rm a}}T^{5/2}_{\rm b}z^{6}-4T^{3/2}_{\rm a}\nonumber\\&&T^{5/2}_{\rm b}z^{6}+2T^{3}_{\rm b}z^{6}-6T_{\rm a}T^{3}_{\rm b}z^{6}\nonumber\\&&-4\sqrt{T_{\rm a}}T^{7/2}_{\rm b}z6{6}-T^{4}_{b}z^{6}+2(z\nonumber\\&&+\sqrt{T_{\rm a}}T_{\rm b}z)^{2}(-16+4*(2-2T_{\rm a}\nonumber\\&&+T^{2}_{\rm a}-2T_{\rm b}+T^{2}_{\rm b})z^{2}+(\sqrt{T_{\rm a}}\nonumber\\&&+\sqrt{T_{b}})^{4}z^{4}   )\rm cos\emph L(\pi+2\theta)-4(z\nonumber\\&&+\sqrt{\emph T_{\rm a}\emph T_{\rm b}}z)^{4}\rm cos2\emph L(\pi\nonumber\\&&+2\theta)] \} \Big/ [(16-4(2-2\emph T_{\rm a}\nonumber\\&&+\emph T^{2}_{\rm a}-2\emph T_{\rm b}+\emph T^{2}_{\rm b})z^{2}+(\sqrt{\emph T_{\rm a}}\nonumber\\&&+\sqrt{\emph T_{\rm b}})^{4}z^{4}+8(z+\sqrt{T_{\rm a}\emph T_{\rm b}}z)^{2}\rm \nonumber\\&&cos2\emph L\theta  )^{1/2}(16  -4(2-2\emph T_{a}+\emph T^{2}_{\rm a}\nonumber\\&&-2\emph T_{\rm b}+\emph T^{2}_{\rm b})\emph z^{2}+(\sqrt{\emph T_{\rm a}}+\sqrt{\emph T_{\rm b}})^{4}\nonumber\\&&z^{4}+8(z+\sqrt{T_{\rm a}T_{\rm b}}z)^{2}\nonumber\\&&\rm  \rm cos\emph L(\pi+2\theta))^{2}].
		\end{eqnarray}

	\bibliography{apssamp}  
\end{document}